\documentclass{article}

\usepackage[left=1in,right=1in,top=1in,bottom=1in]{geometry}
\usepackage[singlespacing]{setspace}
\usepackage{amssymb}
\usepackage{amsfonts}
\usepackage{amsmath}
\usepackage{psfrag}
\usepackage{graphicx}
\usepackage{afterpage}
\usepackage{rotating}
\usepackage{calc}
\usepackage{url}
\usepackage{natbib}

\def\progstart{\singlespacing\begin{center}\begin{minipage}{.95\textwidth}\small\noindent\rule[0pt]{\linewidth}{0.4pt}\vspace{6pt} \\}
\def\progend{\rm\rule[6pt]{\linewidth}{0.4pt} \\ \end{minipage}\end{center}\doublespacing}

\newtheorem{theorem}{Theorem}

\newtheorem{conjecture}[theorem]{Conjecture}

\newtheorem{definition}[theorem]{Definition}
\newtheorem{example}[theorem]{Example}

\newtheorem{lemma}[theorem]{Lemma}

\newenvironment{proof}[1][Proof]{\noindent\textbf{#1.} }{\ \rule{0.5em}{0.5em}}

\begin{document}

\author{Scot Anderson\\ sanderson@southern.edu\\ Southern Adventist University, Tennessee
\and Peter Revesz\\ revesz@cse.unl.edu\\ University of Nebraska-Lincoln}

\title{Efficient Threshold Aggregation of Moving Objects}

\date{}

\maketitle

\begin{abstract}
Calculating aggregation operators of moving point objects, using
time as a continuous variable, presents unique problems when
querying for congestion in a moving and changing (or dynamic) query
space. We present a set of congestion query operators, based on a
threshold value, that estimate the following $5$ aggregation
operations in $d$-dimensions. 1) We call the count of point objects
that intersect the dynamic query space during the query time
interval, the \textsc{CountRange}. 2) We call the Maximum (or
Minimum) congestion in the dynamic query space at any time during
the query time interval, the \textsc{MaxCount} (or
\textsc{MinCount}). 3) We call the sum of time that the dynamic
query space is congested, the \textsc{ThresholdSum}. 4) We call the
number of times that the dynamic query space is congested, the
\textsc{ThresholdCount}. And 5) we call the average length of time
of all the time intervals when the dynamic query space is congested,
the \textsc{ThresholdAverage}. These operators rely on a novel
approach to transforming the problem of selection based on position
to a problem of selection based on a threshold. These operators can
be used to predict concentrations of migrating birds that may carry
disease such as Bird Flu and hence the information may be used to
predict high risk areas. On a smaller scale, those operators are
also applicable to maintaining safety in airplane operations. We
present the theory of our estimation operators and provide
algorithms for exact operators. The implementations of those
operators, and experiments, which include data from more than 7500
queries, indicate that our estimation operators produce fast,
efficient results with error under 5\%.
\end{abstract}

\section{Introduction}

\label{intro:ST}

Safety can often be reduced to to a problem of congestion. The
safety of flight depends on separation of airplanes or more
generally the maximum number of airplanes that a particular airspace
can safely contain, and the maximum number of airplanes that air
traffic controllers (ATC) responsible for directing airplanes can
safely track. When considering epidemics, the presence of a single
animal with Bird Flue does not does not indicate the start of an
epidemic. Instead the presence of a certain number of instances of
the disease indicates a high risk of starting an epidemic, or actual
epidemic conditions. Consequently, we see that congestion often
links to safety and can predict high risk or even dangerous
conditions.

Congestion is defined differently depending on the application.
Hence it is necessary to provide aggregation operators that take a
threshold value as a parameter to define congestion.

In relational databases, \textsc{Max}, \textsc{Min}, \textsc{Count},
\textsc{Sum} and \textsc{Average} form the set of natural
aggregation-operators. Spatiotemporal databases containing moving
objects, based on continuous time, can not apply these operators in
the same way. However, these operators may still function in
interesting ways for moving objects. For example, one can ask how
many moving point objects exist within a moving and changing (or
\emph{dynamic}) rectangular area \emph{at a certain time}, or what
is the maximum distance between two moving points \emph{at certain
times}. Obviously, when we are interested in discrete time
instances, then the moving point object database can be reduced to a
relational database and the above queries can be expressed as simple
\textsc{Count} or \textsc{Max} queries.

Moving object databases naturally suggest new aggregate operators
that have no equivalents in relational databases. For example, one
may ask what is the maximum number of moving-point objects that
exist simultaneously within a dynamic rectangular area at any time
during a time interval $T$? We call this the \textsc{MaxCount} query
(symmetrically we can also find the \textsc{Min-Count}). One may
also ask during what time intervals in $T$ does there exist more
than $M$ moving objects within a rectangular area? We call this the
\textsc{ThresholdRange}. We show that a strong relationship exists
between \textsc{MaxCount} and \textsc{ThresholdRange}, and we show
that \textsc{ThresholdRange} forms the bases for a family of
threshold operators that include: \textsc{ThresholdCount},
\textsc{ThresholdSum}, and \textsc{ThresholdAverage}. A related,
though less complex, operator answers the question: what is the
number of moving objects that exist within or intersect a dynamic
rectangular area at any time instance during interval $T$. We call
this type of query the \textsc{CountRange} query.

We give the following definitions for aggregation operators:

\begin{definition}[Dynamic Query Space]
\label{def:DynamicQuerySpace} Dynamic query space is defined by a
continuous time interval $T$, and a $d$-dimensional space that may
move and change size or shape over the query time interval.
\end{definition}

Throughout this paper we consider the shape of the query space to be
a box or cube.

\begin{definition}[\textsc{MaxCount (MinCount)}]
\label{def:MaxCount} Let $S$ be a set of moving points. Given a
dynamic query space $R$ defined by two moving points $Q_1$ and $Q_2$
as the lower-left and upper-right corners of $R$, and a time
interval $T$, the \textsc{MaxCount} (\emph{Min-Count}) operator
finds the time $t_{\max(\min)}$ and maximum (or minimum) number of
points $M_{\max(\min)}$ in $S$ that $R$ can contain at any time
instance within $T$.
\end{definition}

Throughout this paper we develop the \textsc{MaxCount} operator
because where ever we find a maximum, a minimum can be found
similarly.

\begin{definition}[\textsc{ThresholdRange}]
\label{def:ThresholdRange}Let $S$ be a set of moving points. Given a
dynamic query space $R$ defined by two moving points $Q_1$ and $Q_2$
as the lower-left and upper-right corners of $R$, a time interval
$T$, and a threshold value $M$, the \textsc{ThresholdRange} operator
finds the set of time intervals $T_M$ where the count of objects in
$R$ is larger than $M$.
\end{definition}

\textsc{ThresholdRange} is directly related to \textsc{MaxCount} in
that when $M$ is raised to $M_{\max}$, then \textsc{ThresholdRange}
returns a time interval containing $t_{\max}$ and during this time
interval, the count will be $M_{\max}$.

\begin{definition}[\textsc{ThresholdCount}]
\label{def:ThresholdCount} Given a \textsc{ThresholdRange}, \textsc{%
ThresholdCount} returns the number of time intervals.
\end{definition}

\begin{definition}[\textsc{ThresholdSum}]
\label{def:ThresholdSum} Given a \textsc{ThresholdRange}, \textsc{%
ThresholdSum} returns the total time $T_s$ during which the count is above $%
M $. That is, for each $T_i \in T_M$, \textsc{ThresholdSum} return:
\begin{equation}
T_s = \displaystyle{\sum\limits_i}|T_i|
\end{equation}
where $|T_i|$ means the length of the interval.
\end{definition}

\begin{definition}[\textsc{ThresholdRange}]
\label{def:ThresholdAverage} Given a \textsc{ThresholdRange}, \textsc{%
ThresholdAverage} returns the average length of the intervals in
$T_M$.
\end{definition}

In addition to the threshold aggregation operators, we also use our
bucketing method to implement the \textsc{CountRange} defined as
follows.

\begin{definition}[\textsc{CountRange}]
\label{def:SpatioTemporalRangeCount} Let $S$ be a set of moving
points. Given a dynamic query space $R$ defined by two moving points
$Q_1$ and $Q_2$ as the lower-left and upper-right corners of $R$ and
a time interval $T$, the \textsc{CountRange} query returns the total
number of points that intersect $R$ in $T$.
\end{definition}

Together \textsc{MaxCount (MinCount)} and the threshold operators
form a complete set of threshold aggregation operators comparable to
the aggregation operators given in relational databases.

The following examples use the simple concepts of flying to
demonstrate the use of a few of these threshold aggregation
operators.

\begin{example}
\label{ex:MaxCount}\textrm{Airplanes are commonly modeled as
linearly moving objects with preestablished flight plans. Suppose,
at any time, at most a constant number $M$ of airplanes is allowed
to be in the O'Hare airspace to avoid congestion. Suppose also a new
airplane requests approval of its flight plan for entering the
O'Hare airspace between times $t_a$ and $t_b$. The air traffic
controllers can avoid congestion as follows. If after adding a new
flight plan, the \textsc{MaxCount} between $t_a$ and $t_b$ is still
less than $M$, then they can approve the flight. Otherwise, they
need to find some alternative path, and check it again against the
database. }

\textrm{Air traffic controllers try to direct airplanes as linearly
moving objects for fuel efficiency, among other reasons. If they
recognize a developing congestion too late, then they often must
direct the airplane to fly in circles until the congestion has
cleared. That solution wastes fuel. On the other hand, if they
recognize the developing congestion early, then they can often
simply tell the airplane to change its speed, which saves fuel.
Therefore, it is important to identify congestions as early as
possible. We may identify congestions by using a \textsc{MaxCount}
query where a moving box around the airplane and a time interval
$[t_{a},t_{b}]$ define the query. If the \textsc{MaxCount} predicts
congestion, then the airplane's speed can be adjusted early in the
flight. }
\end{example}

\begin{example}
\label{ex:ThresholdCount}\textrm{Suppose we want to alert pilots if
their current flight path takes them through at least one congested
region. }

\textrm{\emph{Traffic Alert/Collision Avoidance Systems (TCAS)} is a
system that provides similar functionality. TCASs only provide
alerts for current congestion, not predictive congestion. Although
TCASs were implemented in 1986, we continue to have mid-air
collisions and near misses indicating that the system still needs
improvement. \textsc{ThresholdRange} is a modification of
\textsc{MaxCount} that returns all predicted time intervals on the
flight path where the \textsc{Count} exceeds a given threshold.
Hence using \textsc{ThresholdRange} we can alert a pilot of
predicted congestions where more than $M$ other airplanes will be
within the space $B$ around the airplane. Predicting and avoiding
these areas can significantly reduce the chances of mid-air
collisions. }
\end{example}

\begin{example}
\label{ex:CountRange}\textrm{Suppose we are especially concerned
about a rush-hour period $[t_a,t_b]$ that is particularly stressful
to air traffic controllers. Suppose controllers can direct at most
$M$ airplanes safely. We can determine the number of controllers
needed during the rush-hour time by executing the
\textsc{CountRange} query over the controlled airspace during the
rush-hour and dividing by $M$. By ensuring that a sufficient number
of controllers are present, safety is achieved and controllers are
not over stressed. }
\end{example}

Each of the operators can also be applied to examine different
aspects of congestion with regard to bird migration and hence
disease control. These questions and examples, motivated by research
on \textsc{MaxCount}, led us to explore complex threshold
aggregations and data structures to support them.

The rest of this paper is organized as follows.
Section~\ref{sec:BucketDataStructures} gives some background on the
concepts of point domination, sweeping techniques and then
introduces the data structures used to build buckets. These buckets
can then be used in various indexing algorithms to fit the type of
application used. Section~\ref{sec:DynamicMaxCount} develops the
{\sc MaxCount} estimation algorithm using a running example.
Section~\ref{sec:ThresholdOperators} develops the {\sc
ThresholdRange} algorithm based on {\sc MaxCount} and demonstrates
the relationship that ties {\sc MaxCount} to the remaining threshold
operators. This section also develops algorithms for each of those
operators including {\sc CountRange}.
Section~\ref{sec:ExperimentalResults} gives the experimental results
of the implementation. Section~\ref{sec:RelatedWork} reviews the
related work and Section~\ref{sec:Conclusions} gives conclusions and
future work.

\section{Hyper-Bucket Data Structures}\label{sec:BucketDataStructures}

This section presents an updatable {\em skew-aware} bucket for
indices that models the skewed point distributions in each bucket.
The skew-aware technique allows the index structure to perform
inserts, deletes, and updates in {\em fast constant time} using a
\textsc{HashTable} to store the buckets. Many spatiotemporal
applications, such as tracking clients on a wireless network,
particularly need these fast updates and no other {\sc MaxCount}
presented prior to this can meet that requirement. Because the
buckets are spatially defined, the bucketing technique also easily
adapts to other spatial and spatiotemporal indices such as the
\textsc{R-tree}~\cite{DBLP:conf/sigmod/Guttman84}. Hence the
technique performs well for applications where search operations or
update operations occur more frequently by using an appropriate
index.

Our algorithm uses a sweeping method to evaluate the threshold
aggregation operators similar to previous approaches from
\cite{Chen20041,Revesz20031} and \cite{Anderson20061}. The algorithm
differs in that the sweeping algorithm integrates a skew-aware
density function over the spatial dimensions of the bucket to obtain
the time dependent count function. The density function in the
bucket increases accuracy over methods given in
\citep{Chen20041,Anderson20061} while maintaining the same number of
buckets. This idea is a crucial improvement because we model the
point distribution skew in a bucket, whereas previous methods
adapted to skew by increasing the number of buckets or changing
their shape and contents. We also present a precise algorithm for
evaluating the threshold aggregation operators that requires no
index and runs in $O(N)+ O(n \log n)$ time and $O(n)$ space where
$N$ is the number of points in the database and $n$ is the value of
a {\sc CountRange} query using the same query space and time. Both
the threshold aggregation algorithms and the skew-aware bucket data
structure presented are implemented and analyzed in 3-dimensional
space. We show that the approximation achieves good results while
significantly reducing the running times.

Section~\ref{ssec:buckets} describes the problems related to
creating hyper-buckets (also referred to as just buckets) and a
specific solution for creating $6$-dimensional buckets for
$3$-dimensional linearly moving points.  In all cases, we can extend
our method to $d$-dimensions. Section~\ref{ssec:updates} describes
the method for inserting and deleting a point from a bucket and
shows that updates take constant time. Section~\ref{ssec:structures}
applies two different data structures to contain the buckets suited
for applications where either inserts and deletes or threshold
aggregation queries dominate.

\subsection{Hyper-Bucket Data Structure}\label{ssec:buckets}

\begin{definition}[Hex Representation]
\label{def:hex} Define each 3-dimensional linearly moving point $p$
by parametric linear equations in $t$ as follows:
\begin{equation}
    p=\left\{
    \begin{array}{c}
        p_x ~=~ v_x t ~+~ x_0 \\
        p_y ~=~ v_y t ~+~ y_0 \\
        p_z ~=~ v_z t ~+~ z_0 \\
    \end{array}
    \right.
\end{equation}
where the corresponding {\em hex representation} of $p$ is the tuple
$(v_x,x_0,v_y,y_0,v_z,z_0)$ containing the duals of $p_x$, $p_y$,
and $p_z$. For simplicity we often denote the six-tuple as
$(x_1,...,x_6)$.
\end{definition}

\medskip
Consider a relation $D(x_1,..,x_6)$ that contains the {\em hex
representation} of linearly moving points in $3$ dimensions. Then
$D$ represents a $6$-dimensional {\em static} space. Divide the
space into axis-aligned hyper-rectangles where the $k^{th}$ axis has
$d_k$ divisions. Each hyper-rectangle becomes a bucket containing
moving points whose hex falls inside the hyper-rectangle.

\begin{definition}[Hyper-bucket dimensions]
\label{def:bucketDimensions} Define the dimensions of each bucket
$B_i$ by inequalities of the form:
\begin{equation}
\begin{array}{lcl}
    v_{x,L} \leq v_x < v_{x,U} &\bigwedge& x_{0,L} \leq x_0 < x_{0,U} ~\bigwedge \\
    v_{y,L} \leq v_y < v_{y,U} &\bigwedge& y_{0,L} \leq y_0 < y_{0,U} ~\bigwedge \\
    v_{z,L} \leq v_z < v_{z,U} &\bigwedge& z_{0,L} \leq z_0 < z_{0,U} \\
\end{array}
\end{equation}
where we denote the lower bound as:
\begin{equation}
    (v_{x,L}, x_{0,L},v_{y,L}, y_{0,L}, v_{z,L}, z_{0,L})
\end{equation}
and the upper bound as
\begin{equation}
    (v_{x,U}, x_{0,U},v_{y,U}, y_{0,U}, v_{z,U}, z_{0,U}).
\end{equation}
\end{definition}

Each hyper-rectangle defines the spatial dimensions of a possible
bucket, where only buckets that contain points need be included in
the index. The maximum number of possible buckets is given by
$m=\prod\limits_{k}d_k$.

\begin{definition}[Histograms]
\label{def:histograms} Given a $6$-dimensional rectangle $B_i$,
given by Definition~\ref{def:bucketDimensions}, containing $b_i$
points, build the {\em histograms} $h_{i,1}$,...,$h_{i,6}$ for each
axis using $s$ subdivisions as follows. To create histogram
$h_{i,j}$, divide bucket $B_i$ into $s$ parallel subdivisions along
the $j$th axis, and record separately the number of points within
$B_i$ that fall within each subdivision.
\end{definition}

\begin{example}[Building Histograms]\rm
    \begin{figure}[ht]
        \centering
        \psfrag{Y}{$X_0$}
        \psfrag{X}{$V_x$}
        \includegraphics[width=4in]{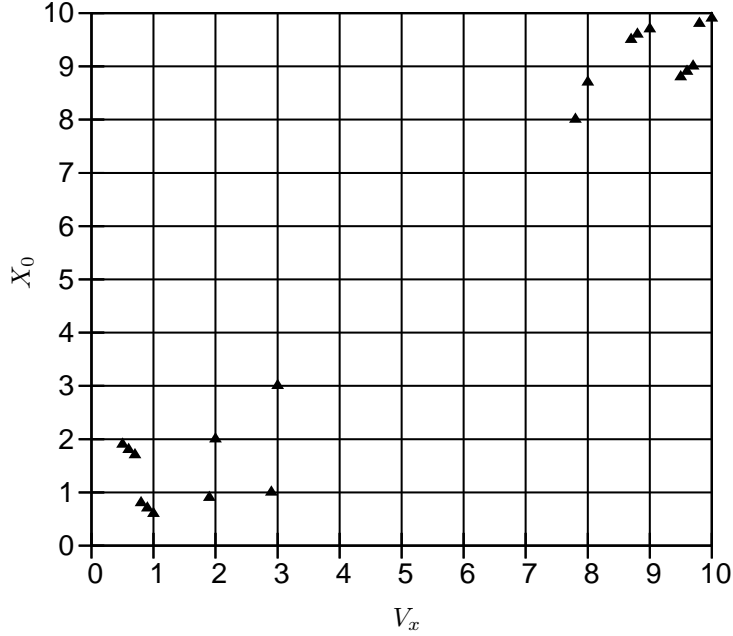}
        \caption{Points projected onto $v_x,x_0$ plane.} \label{fig:points}
    \end{figure}
    Consider a set of 6-dimensional points projected onto the $v_x,x_0$
    plane as shown in Figure~\ref{fig:points}. Assume that the number of
    subdivisions is $s=10$ along both $v_x$ and $x_0$.
    Figure~\ref{fig:vxx0histograms} shows $h_{i,1}$ and $h_{i,2}$. For
    example, the subdivision $0\leq v_x < 1$ contains six points and
    hence the first bar of histogram $h_{i,1}$ rises to level $6$. The
    other values can be determined similarly.

    \begin{figure}[htb]
        \begin{minipage}[t]{3in}
            \begin{center}
            \includegraphics[width=3in]{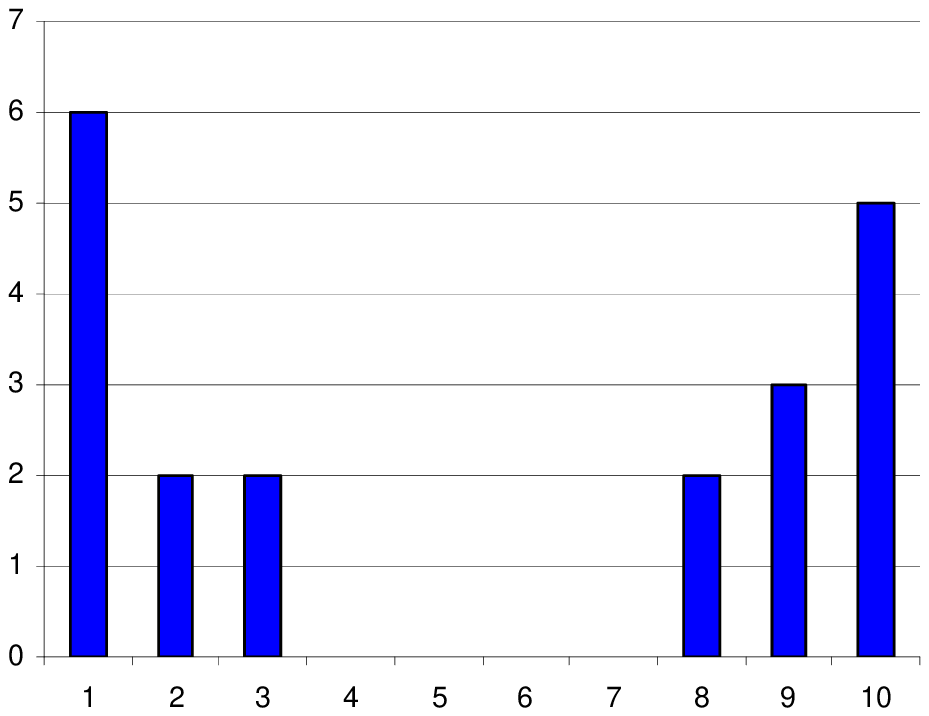}\\
            \mbox{$h_{i,1}$: Points projected onto $v_x$.}
            \end{center}
        \end{minipage}
        \hfill
        \begin{minipage}[t]{3in}
            \begin{center}
            \includegraphics[width=3in]{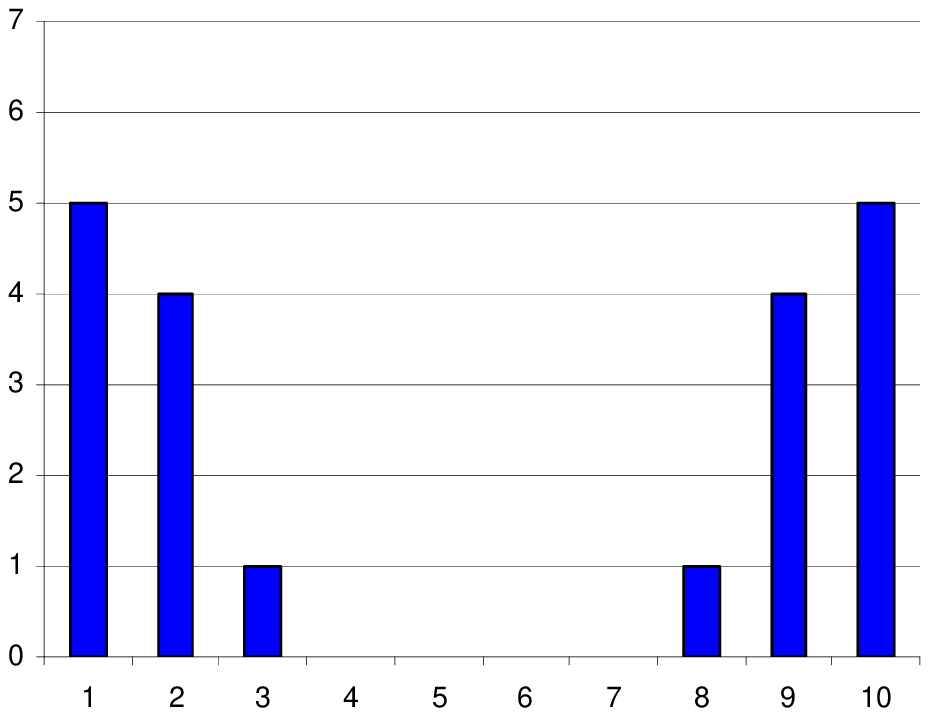}\\
            \mbox{$h_{i,2}$: Points projected onto $x_0$.}
            \end{center}
        \end{minipage}
        \begin{center}
        \end{center}
        \caption{Histogram of Points in 2 Dimensions.}
        \label{fig:vxx0histograms}
    \end{figure}
    \begin{figure}[htb]
            \begin{minipage}[t]{3in}
                \psfrag{A}{$x_0$} \psfrag{B}{$v_x$}
                \includegraphics[width=3in]{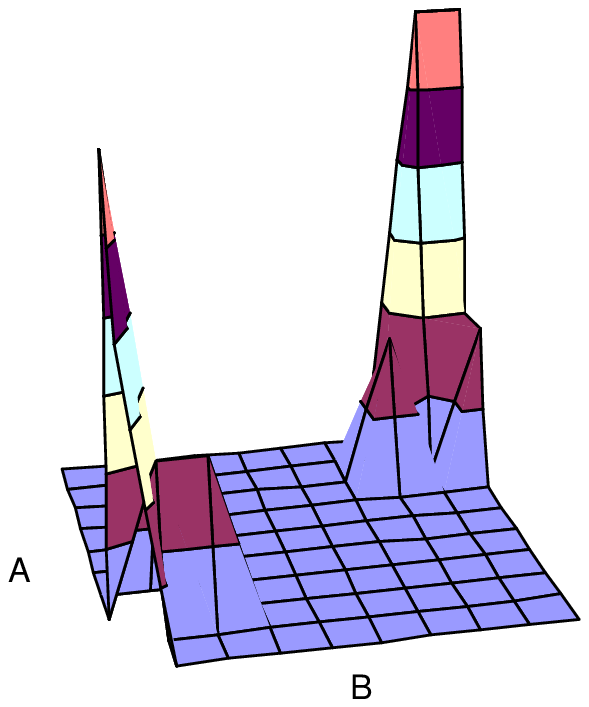}
            \end{minipage}
            \hfill
            \begin{minipage}[t]{3in}
                \psfrag{A}{$x_0$} \psfrag{B}{$v_x$}
                \includegraphics[width=3in]{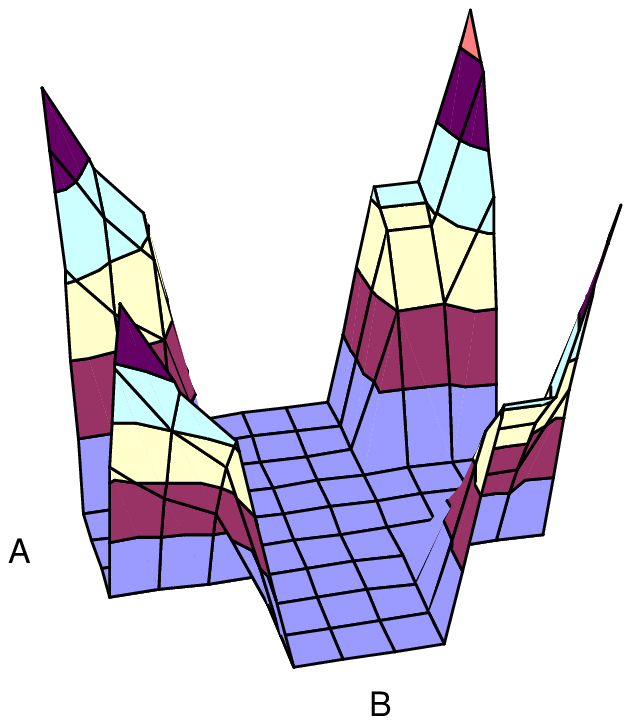}
            \end{minipage}
            \caption{2D Distribution Functions}
            \label{fig:vx2d}
    \end{figure}

    Histograms tell much about the distribution of the points in a
    bucket but they introduce some ambiguity. For example, the
    histograms in Figure~\ref{fig:vxx0histograms} match both of the
    $2d$-distributions in Figure~\ref{fig:vx2d}.
\end{example}
\bigskip

\begin{definition}[Axis Trend Function]
\label{def:trendfunctions}%
The {\em axis trend function} $f_{i,j}(x_j)$ is some polynomial
function for bucket $B_i$ and axis $j$ such that the following hold:
\begin{enumerate}
    \item $f_{i,j} \geq 0$ over $B_i$.
    \item $f'_{i,j}$, the derivative $f_{i,j}$, does not change sign over the valid range.
\end{enumerate}
The {\em bucket trend function} $f_i$ for bucket $B_i$ is the
following:
    \begin{equation}
        \label{eq:bucketdensity}
        f_i=\prod_j f_{i,j}
    \end{equation}
\end{definition}

Condition 1 ensures that the bucket trend function, built from the
axis trend functions, does not contain a negative probability
region. Condition 2 requires that the bucket density increase,
decrease, or remain constant when considering any single axis. This
condition avoids the ambiguity demonstrated in
Figures~\ref{fig:vxx0histograms} and \ref{fig:vx2d} by giving a
polynomial that approximates the density change correctly. We show
this in the following Lemma.

\begin{lemma}
\label{lem:distributionindependence} Given a bucket $B_{i}$ with
bucket trend functions $f_{i,j}$, let $r_{1}$ and $r_{2}$ be
identically sized regions in bucket $B_{i}$. If the density in
$B_{i}$ along each axis monotonically increases from $r_{1}$ to
$r_{2}
$ the following holds:%
\begin{equation}
\int_{r_{2}}f_{i}~d\phi \geq \int_{r_{1}}f_{i}~d\phi
\end{equation}
\end{lemma}

\begin{proof}
Increasing densities from $r_{1}$ to $r_{2}$ translates into
histograms that also increase from $r_{1}$ in the direction of
$r_{2}$ along each axis. The translation from histograms to the axis
trend functions gives the following conditions:
\begin{equation}
    f_{i,j}(x_{2,j})\geq f_{i,j}( x_{1,j})
\end{equation}
where $x_{1,j}$ and $x_{2,j}$ are the $j^{th}$ coordinates of the
points in $r_{1}$ and $r_{2}$ respectively, and are located the same
distance from the $j^{th}$ coordinates of the lower bounds of $r_1$
and $r_2$ respectively. Since this constraint holds for each $j$ and
$f_{i,j}\geq 0$ we have:
\begin{equation}
    f_{i}(x_2)\geq f_{i}(x_1)
\end{equation}
Hence by the properties of integration we conclude
\begin{equation}
    \int_{r_{2}}f_{i}~d\phi \geq \int_{r_{1}}f_{i}~d\phi
\end{equation}
\end{proof}

Definition~\ref{def:trendfunctions} allows a whole class of
polynomial functions, and Lemma~\ref{lem:distributionindependence}
applies to each member of that class. However, in the following, we
use a particular polynomial function derived from the product of
linear functions, which are obtained by using the least squares
method for each histogram.

\begin{definition}[Normalized Trend Functions]
\label{def:NormalizedTrendFunction} Let $n$ be the number of points
in the database, $b_i$ the number of points in bucket $B_i$, and
$f_i$ be given by Equation~(\ref{eq:bucketdensity}). The {\em
normalized trend function} $F_i$ for bucket $B_i$ is:
\begin{equation}
    F_{i} = \frac{b_i f_{i}}
    {
        n \mathop{\displaystyle\int}\limits_{B_i}^{~}
        f_{i}~d\phi
    }
    \label{eq:NormalizedSurface}
\end{equation}
and the {\em percentage of points} in bucket $B_i$ is:
\begin{equation}
        p = \mathop{\displaystyle\int}\limits_{B_i} F_i~d\phi.
\label{eq:percentagepoints}
\end{equation}
\end{definition}

With this definition we can calculate the number of points in $O(1)$
time using the following simple lemma.

\begin{lemma}
\label{lem:ConstRunningTimeForBucket} Let $B_i$ be a bucket, $n$ the
number of points in the databases, and $p$ be given by
Definition~\ref{def:NormalizedTrendFunction}. Then $np$ is the
number of points in bucket $B_i$ and $np$ is calculated in $O(1)$
time.
\end{lemma}
\begin{proof}
By Equation~(\ref{eq:NormalizedSurface}) and
(\ref{eq:percentagepoints}) we have:
\begin{equation}
\begin{array}{ccl}
    n p & = &n \mathop{\displaystyle\int}\limits_{B_i} F_i~d\phi \vspace{6pt}\\
        & = &n  \mathop{\displaystyle\int}\limits_{B_i} \displaystyle{\frac{b_i}{n}} \frac{f_{i}}{\mathop{\displaystyle\int}_{B_i} f_i~d\phi}~d\phi \vspace{6pt}\\
        & = &n  \displaystyle{\frac{b_i}{n}} \cdot \frac{\mathop{\displaystyle\int}_{B_i} f_{i}~d\phi}{\mathop{\displaystyle\int}_{B_i} f_{i}~d\phi} \vspace{6pt}\\
        & = &b_i. \\
\end{array}
\end{equation}
Clearly the above calculations take only $O(1)$ time.
\end{proof}

Using the above definitions we can now define the bucket data
structure used throughout the rest of this paper.

\begin{definition}[Skew Aware Buckets]\label{def:bucketsN}%
A bucket is a hyper-rectangle with dimensions given by
Definition~\ref{def:bucketDimensions} and that maintains histograms
given by Definition~\ref{def:histograms}, additional data for the
least squares method, and the normalized trend function given by
Definition~\ref{def:NormalizedTrendFunction}. Throughout the rest of
this paper we refer to these as buckets.
\end{definition}

\subsection{Inserts and Deletes}\label{ssec:updates}

We can maintain the bucket (and hence the index) while deleting or
inserting a point for any bucket $B_i$ by recalculating the trend
function $F_i$ for the bucket.

\begin{lemma}\label{lem:ConstantUpdates}
Insertion and deletion of a moving point can be done in $O(1)$ time.
\end{lemma}

\begin{proof}
When we insert or delete a point, we need to update the histograms
and the normalized trend function. Let the point to insert/delete be
$P_a$ represented using the hex representation as
$(a_0,a_1,a_2,a_3,a_4,a_5)$, let $d_j$, for $0 \leq j \leq 5$ be the
bucket width in the $j^{th}$, and let $s$ be the number of
subdivisions in each histogram. The concatenation of $id_0, \ldots,
id_5$ gives the $ID_i$ of bucket $i$ to insert (or delete) $P_a$
into where each $id_l$ and $0 \le l \le 5$ is defined by:
\begin{equation}
    id_l = \left\lfloor \frac{a_l}{d_l} \right\rfloor.
\end{equation}
The calculation of $ID_i$ and retrieving bucket $B_i$ takes $O(1)$
time using a \textsc{HashTable}.

Let $hw_{i,j}$ be the histogram-division width for the $j^{th}$
calculated as $hw_{i,j} = \left\lceil \frac{d_j}{s} \right\rceil$.
Then $p$ is projected onto each dimension to determine which
division of the histogram to update. For the $j^{th}$ dimension the
$k^{th}$ division of histogram $h_{i,j}$ is given as follows:
\begin{equation}
    k(j) = \left\lfloor \frac{a_j - id_j*d_j}{hw_k} \right\rfloor
\end{equation}
Let $h_{i,j,k}$ be the histogram division to update for each
histogram. Update $h_{i,j,k}$ and the sums $\displaystyle{\sum}y_i$,
and $\displaystyle{\sum}x_i y_{i}$ from the normal equations in the
least squares method. $N$, $\displaystyle{\sum}x_i$ and
$\displaystyle{\sum}x_{i}^{2}$ from the normal equations do not need
updating since the number of histogram divisions $s$ is fixed within
the database.

We can now recalculate each $f_{i,j}$ in constant time by solving
the $2 \times 3$ matrix corresponding to the normal equations of the
least squares method for each histogram. For each $f_{i,j}$
calculate the endpoints to determine the required shift amount
(Definition~\ref{def:trendfunctions}, property 1) and calculate
$f_i$ from Equation~(\ref{eq:bucketdensity}). Now we calculate $F_i$
using Equation~(\ref{def:NormalizedTrendFunction}). Each of these
steps depends only on the dimension of the database. Hence for any
fixed dimension we can rebuild the normalized trend function $F_i$
in $O(1)$ time.
\end{proof}

\subsection{Index Data Structures}\label{ssec:structures}

There is no need to create a bucket unless it contains at least one
point. We consider two classes of data structures for organizing the
buckets: \textsc{HashTables} and \textsc{Trees}.

For databases where inserts and deletes are the most common
operation, the \textsc{HashTable} approach allows these operations
to run in constant time. However, the {\sc MaxCount} operation will
require an enumeration of all the buckets and thus at least a
running time of $O(B)$. As long as the number of buckets is
reasonable, this approach works well.

For databases where {\sc MaxCount} is the most common operation, we
may use an \textsc{R-tree} structure
\citep{DBLP:conf/sigmod/Guttman84,BKS+90} where the elements to be
inserted are the buckets. This approach speeds up the {\sc MaxCount}
query to $O(\log|B| + R)$ where $R$ is the number of buckets needed
to calculate the query. The insert and delete costs for these
\textsc{R-trees} are $O(\log|B|)$, because buckets do not overlap.

Since buckets do not change shape, the database is decomposable and
allows each type of aggregation to be calculated from simultaneous
executions on subspaces of the index space. We discuss the method
and ramifications of this capability at the end of Section
\ref{sec:ExactMaxCount}.

\section{Dynamic \textsc{MaxCount}}\label{sec:DynamicMaxCount}

Section~\ref{ssec:PointDomination} reviews point domination in
higher dimensions. Section~\ref{ssec:IntegratingBuckets} examines
finding the percentage of points in a bucket that are in the query
space as a function of time. Section~\ref{ssec:MaxCountAlgorithm}
puts the two previous sections together to create the dynamic {\sc
MaxCount} algorithm for $d$-dimensions.

\subsection{Point Domination in 6-Dimensional Space}\label{ssec:PointDomination}

Let $B$ be the set of 6-dimensional hyper-buckets in the input where
each hyper-bucket $B_i$ has an associated normalized trend function
$F_i$ as in Definition~\ref{def:NormalizedTrendFunction}. Let the
vertices of $B_i$ be denoted $v_{i,j}$ where $1 \leq j \leq 64$,
because there are $2^6$ corner vertices to a 6-dimensional
hyper-cube.

\begin{definition}[Point Domination]\label{def:pointdomination}
    Given two linearly moving points in three dimensions
    \begin{equation}
            P(t)=\left\{
        \begin{array}{l}
            p_{x}=x_1 t + x_2 \\
            p_{y}=x_3 t + x_4 \\
            p_{z}=x_5 t + x_6
        \end{array}
        \right.   
        \quad {\rm and} \quad
        Q(t)=\left\{
        \begin{array}{l}
            q_{x}=v_{x}t+x_{0} \\
            q_{y}=v_{y}t+y_{0} \\
            q_{z}=v_{z}t+z_{0}
        \end{array}
        \right.
    \end{equation}
    $Q(t)$ dominates $P(t)$ if and only if the following holds:
    \begin{equation}
        (p_x < q_x) \quad \wedge \quad (p_y < q_y) \quad \wedge \quad (p_z < q_z).
    \end{equation}
\end{definition}

The previous definition takes 6-dimensional points defined in
Definition~\ref{def:hex} and places them into three inequalities of
the form $x_2 < -t(x_1-v_x) + x_0$. Each inequality defines a region
below a line with slope $-t$.

\begin{figure}[ht]
    \centering
    \psfrag{X-View}{$X-$view}
    \psfrag{Y-View}{$Y-$view}
    \psfrag{Z-View}{$Z-$view}
    \psfrag{Q2x}{$Q_{2x}$}
    \psfrag{Q1x}{$Q_{1x}$}
    \psfrag{Q2y}{$Q_{2y}$}
    \psfrag{Q1y}{$Q_{1y}$}
    \psfrag{Q2z}{$Q_{2z}$}
    \psfrag{Q1z}{$Q_{1z}$}
    \psfrag{lx2}{$l_{x2}$}
    \psfrag{lx1}{$l_{x1}$}
    \psfrag{ly2}{$l_{y2}$}
    \psfrag{ly1}{$l_{y1}$}
    \psfrag{lz2}{$l_{z2}$}
    \psfrag{lz1}{$l_{z1}$}
    \psfrag{Position}{Position}
    \psfrag{Velocity}{Velocity}
    \includegraphics[width=5.9in]{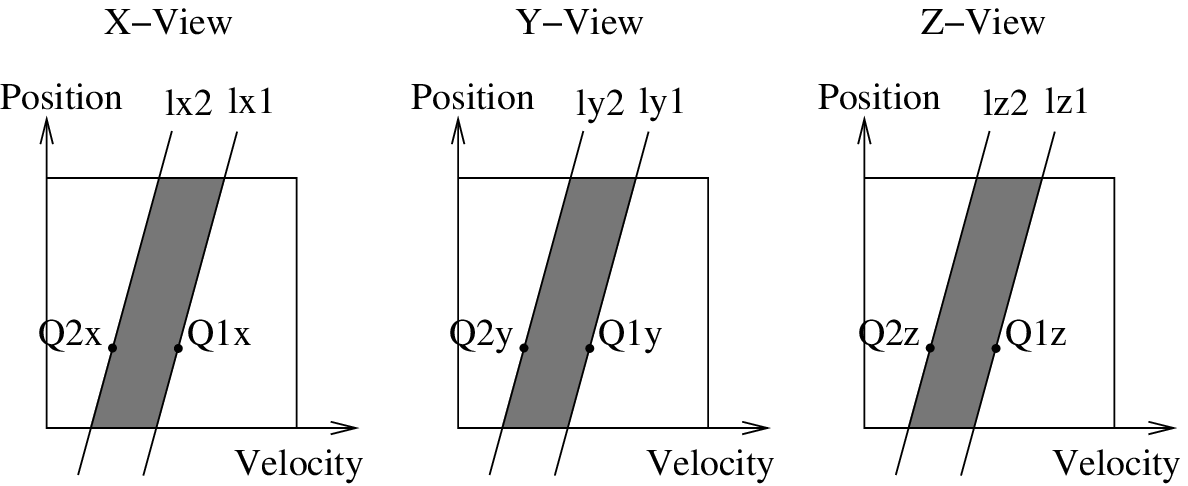}\\
    \caption{Views.}\label{fig:views}
\end{figure}

\begin{definition}[$x$-view, $y$-view and $z$-view projections]\label{def:views}
Projecting the inequalities from
Definition~\ref{def:pointdomination} onto their respective dual
planes allows a visualization in three 2-dimensional planes. Define
these three projections as the $x-$view, $y-$view and $z-$view
respectively. Because the time $-t$ defines the slopes of each line,
all views contain lines with identical slopes. (See
Figure~\ref{fig:views})
\end{definition}

\begin{definition}[Query Space]\label{def:queryspace}
Given two moving query points $Q_1(t)$ and $Q_2(t)$ and lines
$l_{x1}$, $l_{x2}$, $l_{y1}$, $l_{y2}$, $l_{z1}$, $l_{z2}$ crossing
them in their respective hexes with slopes $-t$, the intersection of
the bands formed by the area between $l_{x1}$ and $l_{x2}$, $l_{y1}$
and $l_{y2}$, and $l_{z1}$ and $l_{z2}$ in the 6-dimensional space
forms a hyper-tunnel that defines the {\em query space} as shown in
Figure~\ref{fig:views}.
\end{definition}

We can now visualize the query in space and time as the {\em query
space} sweeping through a bucket as the slopes of the lines change
with time. Using the above, it is now easy to prove the following
lemma.

\begin{lemma}
At any time $t$, the moving points whose hex-representation lies
below (or above) $l_{x1},l_{y1}$ and $l_{z1}$ in their respective
views are exactly those points that lie below (or above) $Q_{1}$ in
the original 3-dimensional plane.
\end{lemma}

\begin{proof}
Let $Q_{x}(t)=v_{x}t+x_{0}$ where $v_{x}$ and $x_{0}$ are constants
and consider any $x$ component of a point $P_{x}(t)=x_1 t + x_2$
that lies below $Q$ on the $x$-axis. Then
\begin{eqnarray}
    x_1 t + x_2 &<& v_{x} t + x_{0} \\
    x_2         &<& -t (x_1 - v_{x}) + x_{0}
\end{eqnarray}
Obviously, at any time $t$ these are the points below the line $x_2
= -t(x_1 - v_{x}) + x_{0}$, which has a slope of $-t$ and goes
through $( v_{x},x_{0})$. This representation is the dual of point
$Q_{x}$. By Definition \ref{def:queryspace}, this is exactly the
line $l_{x1}$. We can prove similarly that the points with duals
above $l_{x1}$ are above $Q_{1}$ at any time $t$. The proof that
points whose hex-representations are above or below $l_{y1},$ and
$l_{z1}$ are exactly those points that lie above or below $Q_{1}$ is
similar to the proof for points above or below $l_{x1}$. By
Definition~\ref{def:pointdomination}, we conclude that the points
dominated by $Q_{1}$ in the dual space are those points that are
below $l_{x1}, l_{y1}$, and $l_{z1}$ in the $x$-view, $y-$view, and
$z$-view, respectively. Similarly, we conclude that the points that
dominate $Q_1$ in the dual space are those points that are above
$l_{x1},~l_{y1}$, and $l_{z1}$ in the $x$-view, $y-$view, and
$z$-view, respectively.
\end{proof}

Throughout the examples in this chapter, we use the points shown in
Figures~\ref{fig:Points} and \ref{fig:ExPointsProjected} to
demonstrate the evaluation of a {\sc MaxCount} query. We begin by
creating the index.

\begin{example}[Creating the Index]\label{ex:BuildIndex}\rm
    \begin{figure}[htb]
        \centering
        \includegraphics[scale=1]{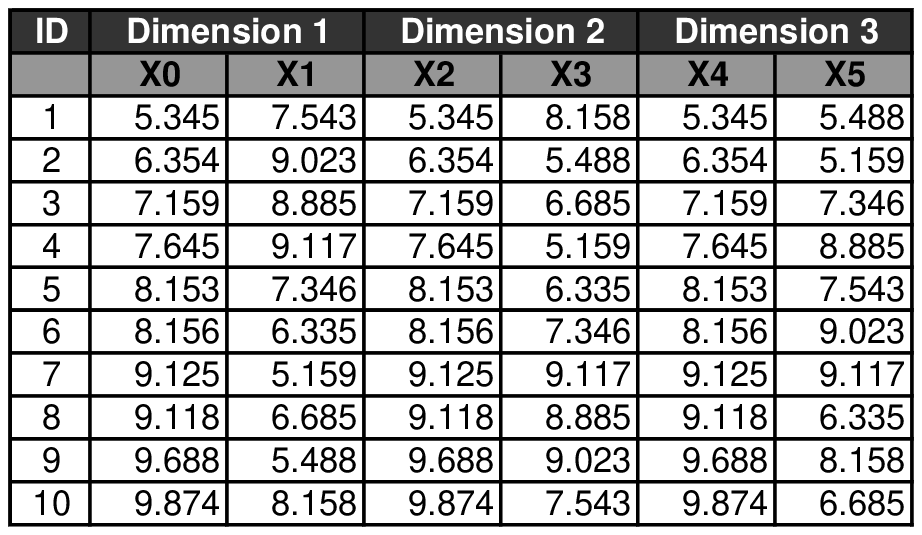}
        \caption{Example points.}
        \label{fig:Points}
    \end{figure}

    Consider a relation that contains the $6$-dimensional space 10
    units $(0 \ldots 10)$ in each dimension. If we break this up
    into buckets that are $5$ units long in each dimension, we have
    $2^{6}$ buckets. Although these divisions make a space with $64$
    buckets, all the points are contained in a single bucket whose
    index is $(2,2,2,2,2,2)$. All the points listed in Figure
    \ref{fig:Points} have the same velocities for each dual plane.
    Notice the columns for $x_1$, $x_3$, and $x_5$ all have the same
    values in different orders. The projection of the points onto
    the $3$ dual planes shown in Figure~\ref{fig:ExPointsProjected}
    does not immediately show this organization. Projecting the
    points for any view in Figure~\ref{fig:HistogramVP} onto each
    axis and creating histograms with $5$ divisions gives the
    histograms for the Velocity and Position axes shown in
    Figure~\ref{fig:HistogramVP}.
    \begin{figure}[h]
        \centering
        \begin{minipage}[t]{2in}
            \begin{center}
                \includegraphics[width=2in]{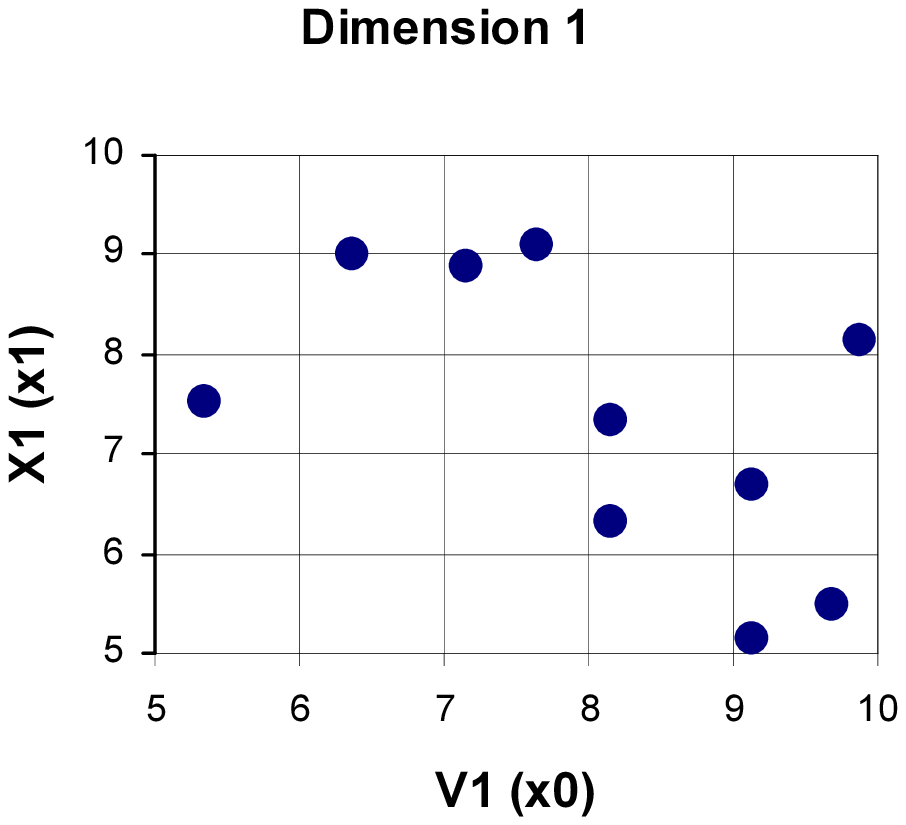} \\
                (a)
            \end{center}
        \end{minipage}
        \begin{minipage}[t]{2in}
            \begin{center}
                \includegraphics[width=2in]{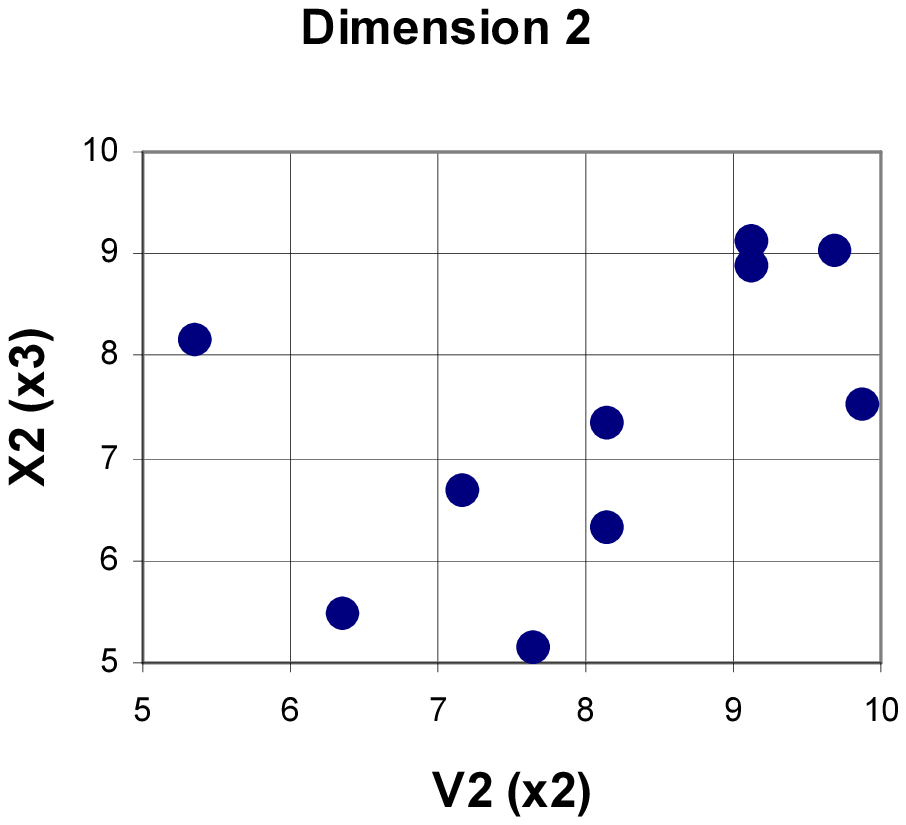} \\
                (b)
            \end{center}
        \end{minipage}
        \begin{minipage}[t]{2in}
            \begin{center}
                \includegraphics[width=2in]{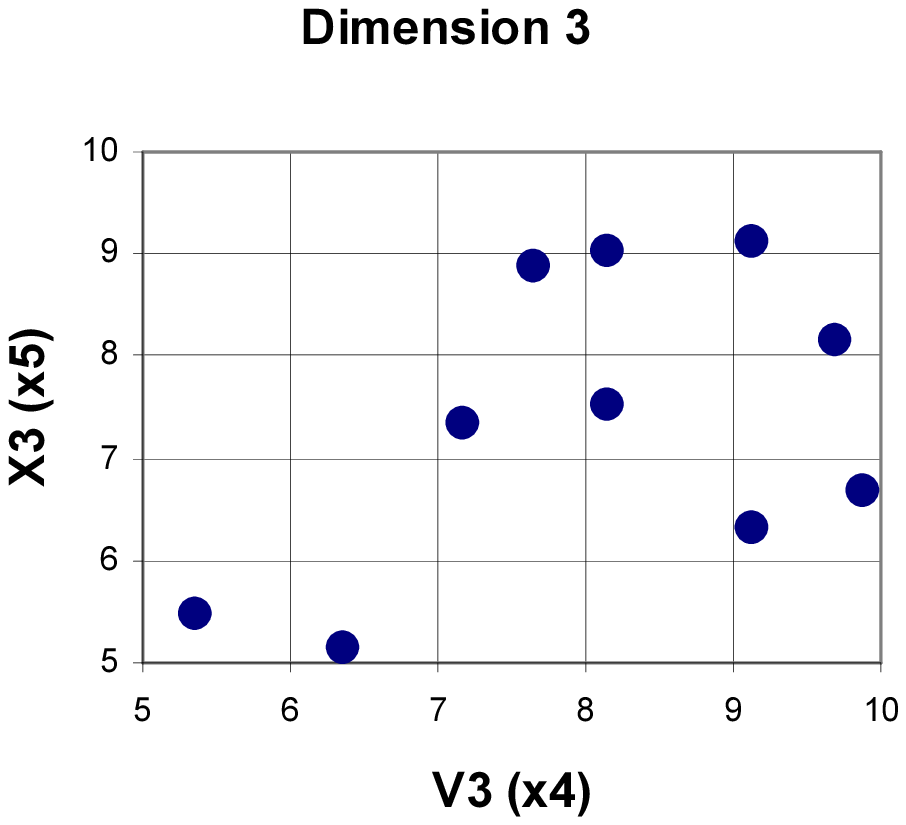} \\
                (c)
            \end{center}
        \end{minipage}
    \caption{Points projected onto (a) $X$-view, (b) $Y$-view, and (c) $Z$-view.}
    \label{fig:ExPointsProjected}
    \end{figure}
    \begin{figure}[h]
        \centering
        \begin{minipage}[t]{3in}
            \begin{center}
                \includegraphics[width=3in]{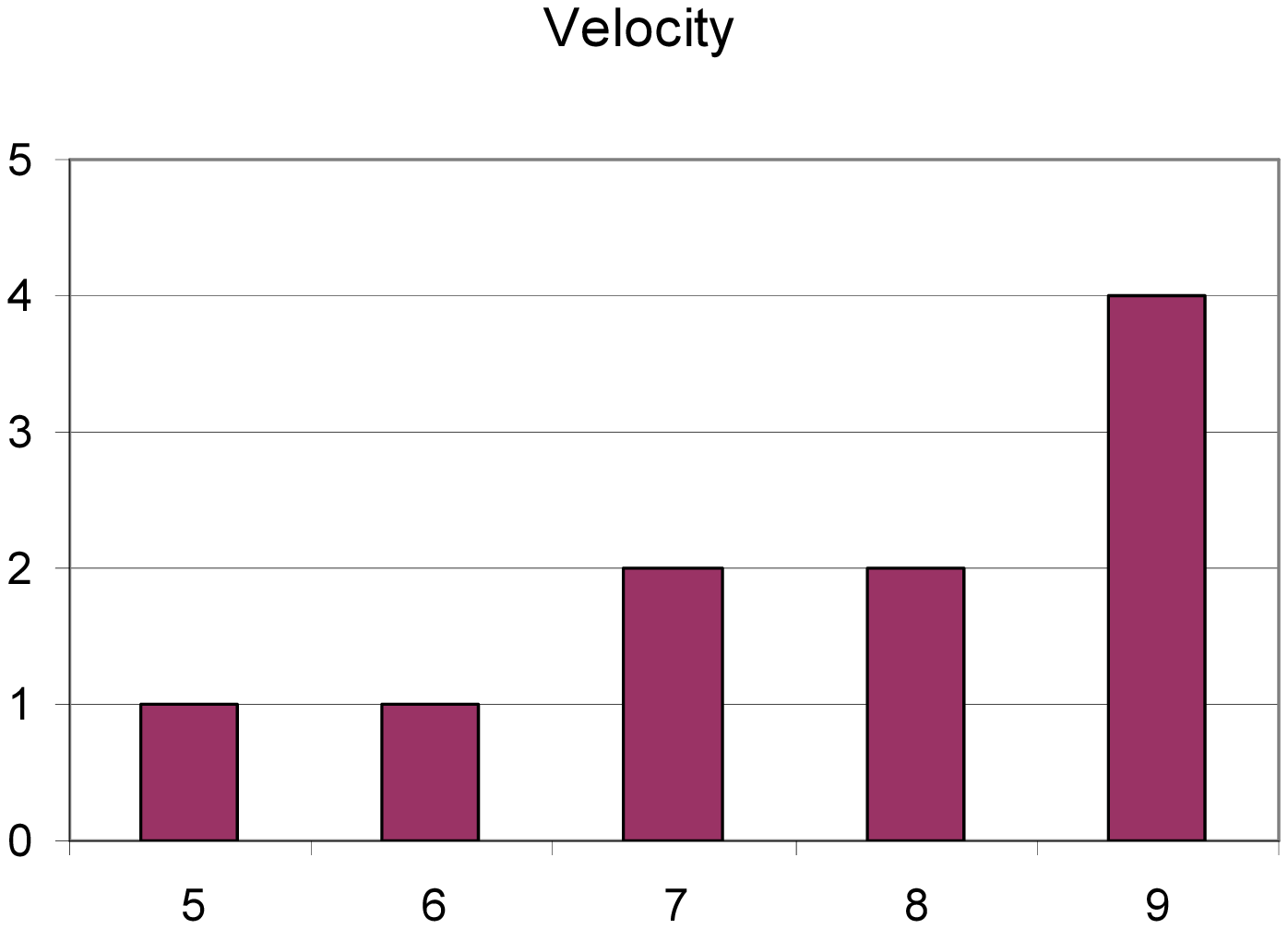} \\
                (a) Velocity.
            \end{center}
        \end{minipage}
        \begin{minipage}[t]{3in}
            \begin{center}
                \includegraphics[width=3in]{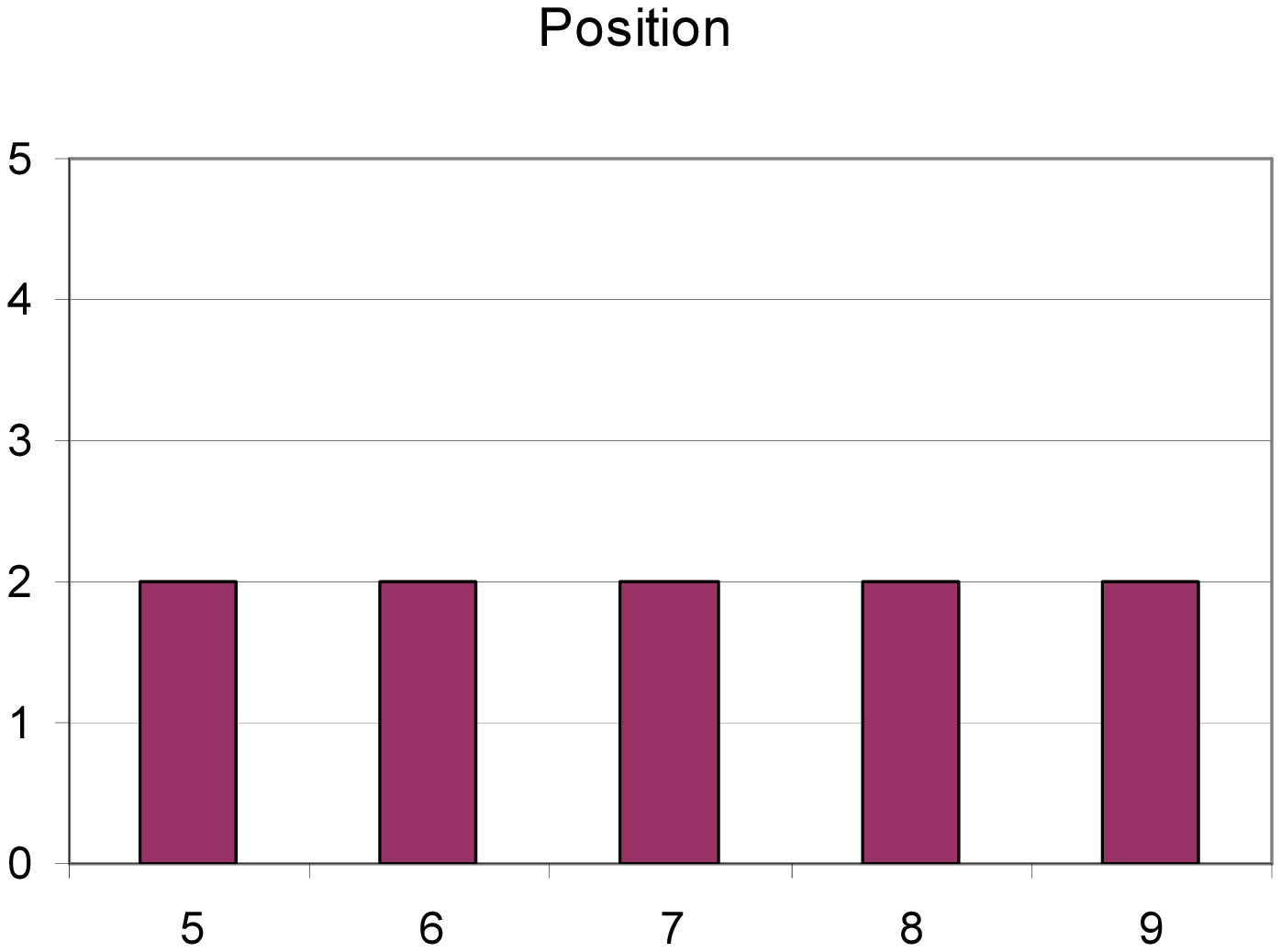} \\
                (b) Position.
            \end{center}
        \end{minipage}
        \caption{Position and velocity histograms, identical for each view.}
        \label{fig:HistogramVP}
    \end{figure}
    Hence, each velocity dimension has the same histogram. Similarly each
    position dimension has the same histogram. To create these
    histograms each point is projected onto the axis. For example point
    $1$ projected onto the $x_1$ axis is given as:
    \begin{equation}
    ~5.345,7.543,5.345,8.158,5.345,5.488\rightarrow5.345.
    \end{equation}
    Calculate the widths of the histograms as:
    \begin{equation}
        Histogram\_Width =(10-5)/5 =1
    \end{equation}
    We determine the histogram for each point by looping through the
    points and calculating the following:
    \begin{equation}
        division=\left\lfloor((point-lowerbound)/Histogram\_Width)\right\rfloor
    \end{equation}
    For example the lowest and highest points in velocity would be added
    to the division calculated as
    $\left\lfloor \left(  5.84-5\right)  /1\right\rfloor = 0$ and
    $\left\lfloor (9.468-5)/1\right\rfloor = 4$.

    The histograms translate into a set of points for each view given
    as:
    \begin{eqnarray}
        Velocity =\{(0,1),(1,1),(2,2),(3,2),(4,4)\}\label{pt:Vel} \\
        Position =\{(0,2),(1,2),(2,2),(3,2),(4,2)\}\label{pt:Pos}
    \end{eqnarray}
    Before applying the least squares method each division number
    must be translated back into the bucket. Translation is done
    using the following code fragment:
    \medskip

    \progstart \vspace{-18pt}
    \begin{tabbing}
    \hspace{.25in}\= \kill
    \textbf{for} $i \leftarrow 0$ \textbf{to} \emph{number\_of\_divisions} $-1$\\
    \> $point[i][0]\ \leftarrow i*histogram\_width + lowerbound$ \\
    \> $point[i][1]\ \leftarrow histogram\_value[i]$ \\
    \textbf{end for}
    \end{tabbing}
    \progend

    Translation of the points from (\ref{pt:Vel}) and (\ref{pt:Pos}) gives: The
    histograms for velocity and position in each view are given as:
    \begin{eqnarray}
        Velocity =\{(5,1),(6,1),(7,2),(8,2),(9,4)\} \\
        Position =\{(5,2),(6,2),(7,2),(8,2),(9,2)\}.
    \end{eqnarray}
    Using the least squares method to fit each of these to a line yields
    the following for each velocity and position dimension:
    \begin{align}
        Velocity:~~  &y=0.7x-2.9\label{eq:RawVelocity}\\
        Position:~~  &y=0x+2 \label{eq:RawPosition}.%
    \end{align}
    Evaluating Equations~(\ref{eq:RawVelocity}) and
    (\ref{eq:RawPosition}) at the end points to find the shift value
    for the axis trend function to add to each equation gives:
    \begin{align}
        Velocity:~~  &y(5)=1,~~ y(10)=4.3\\
        Position:~~  &y(5)=y(10)=2.
    \end{align}
    In this case no constant needs to be added to our equation and
    the trend function becomes:
    \begin{equation}
        f_{i}=(0.7x_{0}-2.9)(0x_{1}+2)(0.7x_{2}-2.9)(0x_{3}+2)(0.7x_{4}-2.9)(0x_{5}+2)
    \end{equation}
    Calculating $F_{i}$ from Equation~(\ref{eq:NormalizedSurface}) requires
    integrating $f_i$ over the bucket where
    $\int_{B_{i}}\equiv\int_{5}^{10}...\int_{5}^{10}$ and where
    $d\phi\equiv
    dx_{0}dx_{1}dx_{2}dx_{3} dx_{4}dx_{5}$ gives
    \begin{align}
        \int_{B_{i}}f_{i}d\phi &  =8 \int_{B_{i}}(0.7x_{0}-2.9)(0.7x_{2}-2.9)(0.7x_{4}-2.9)d\phi \nonumber\\
                            &  =1622234.375.
    \end{align}
    Since all the points reside in a single bucket, $b_{i}=n$, the
    constant $c$ is given by $c=1/1622234.375 \approx 6.164\times10^{-7}$. Then
    $F_{i}$ is given by
    \begin{align}
        F_{i}  &  \approx c~(0.7x_{0}-2.9)(0x_{1}+2)(0.7x_{2}-2.9)(0x_{3}+2)(0.7x_{4}-2.9)(0x_{5}+2)\label{eq:Fi}\nonumber\\
            &  =8c(0.7x_{0}-2.9)(.7x_{2}-2.9)(.7x_{4}-2.9)
    \end{align}
    So far we have calculated the normalized trend function $F_{i}$
    for just one bucket. This calculation finishes the bucket
    creation process, and the index contains this single bucket
    defined by the points $lowerbound=(5,5,5,5,5,5)$ and
    $upperbound=(10,10,10,10,10,10)$.
\end{example}

\subsection{Approximating the Number of Points in a Bucket}\label{ssec:IntegratingBuckets}

\begin{figure}[ht]
    \centering
    \includegraphics[width=4.25in]{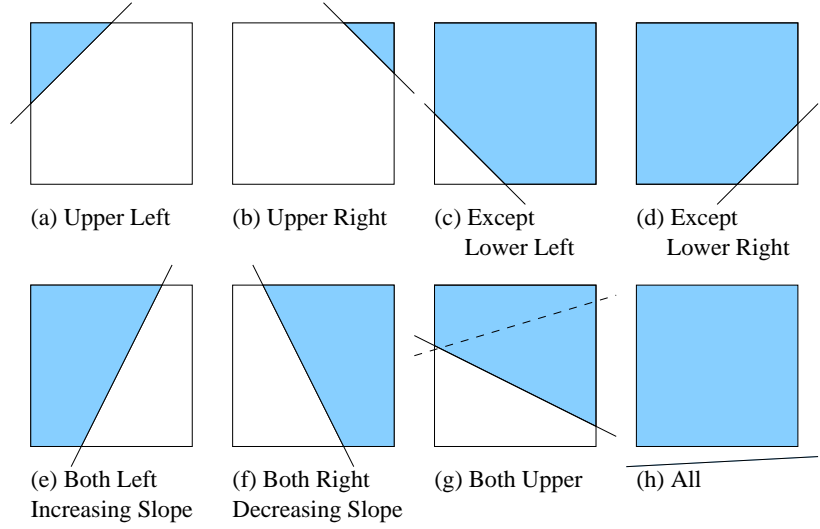}\\
    \caption{Sweep algorithm cases.}
    \label{fig:cases}
\end{figure}

As a line through a query point sweeps across a bucket, the points
in the bucket that dominate the query point are approximated by the
integral over the region above the line. In each of the three views
the query space intersects the plane giving the cases shown in
Figure~\ref{fig:cases}.

\begin{definition}[Percentage Function]\label{def:percentagefunction}
Integrating over the region above the line gives an approximation of
the percentage of points in the query space. We define the
percentage function given as:
\begin{equation}\label{eq:percentofbucket}
    p=\int\limits_{r_1} F_i~d\phi
\end{equation}
where $r_1$ is the region of the bucket in the query space. If two
lines go through the same bucket we have the smaller region $r_2$
subtracted from the larger region $r_1$ as follows.
\begin{equation}\label{eq:percentofbucket2}
    \triangle p=\int\limits_{r_1} F_i~d\phi - \int\limits_{r_2} F_i~d\phi.
\end{equation}
Here, regions $r_1$ and $r_2$ correspond to regions above $Q_1$ and
$Q_2$ in Figure~\ref{fig:views}, respectively.
Lemma~\ref{lem:ConstRunningTimeForBucket} showed that finding the
number of points in the bucket requires multiplying
Equation~(\ref{eq:percentofbucket}) or (\ref{eq:percentofbucket2})
by $n$.
\end{definition}

For each case shown in Figure~\ref{fig:cases}, we describe the
function that results from integration in one view. To extend the
result to any number of views, we take the result from the last view
and integrate it in the next view. If the region below the line were
desired, $p_{lower}=\frac{b_i}{n}-p$ gives the percentage of points
below the line.

For cases (a) -- (h) below, let $Q=(x_{1,q},x_{2,q},...,x_{6,q})$.
For the $x$-view, let the lower left corner vertex be
$(x_{1,l},x_{2,l})$ and the upper right corner vertex be
$(x_{1,u},x_{2,u})$. In addition each line denoted $l$ is given by
$x_2 = -t (x_1 - x_{i,q}) + x_{i+1,q}$ and corresponds to a line
shown in the corresponding case in Figure~\ref{fig:cases}.

\medskip\noindent{\bf Case (a):}
For this case $l$ crosses the bucket at $x_{1,l}$ and $x_{2,u}$. The
integral over the shaded region is given by the following:
\begin{equation}\label{eq:integrala}
    p_a = \int\limits_{x_{1,l}}^{\frac{x_{2,u} - x_{2,q}}{-t} + x_{1,q}}
    \int\limits_{-t(x_1 - x_{1,q}) + x_{2,q}}^{x_{2,u}}
    F_i~dx_2 dx_1
\end{equation}
Notice that the lower bound of the integral over $dx_2$ contains
$x_1$. This dependence within each view does not affect the
integration in the remaining four dimensions. The solution to
Equation~(\ref{eq:integrala})
has the form:
\begin{equation}\label{eq:forma}
    a t^2 + b t + c + \frac{d}{t} + \frac{e}{t^2}.
\end{equation}

\medskip\noindent{\bf Case (b):}
For this case $l$ crosses the bucket at $x_{1,u}$ and $x_{2,u}$. The
integral over the shaded region is given by:
\begin{equation}\label{eq:integralb}
    p_b = \int\limits_{-\frac{(x_{2,u}-x_{2,q})}{t}+x_{1,q}}^{x_{1,u}}\int
    \limits_{-t(x_{1}-x_{1,q})+x_{2,q}}^{x_{2,u}}F_i~dx_{2}dx_{1}.
\end{equation}
The solution
has the form of Equation~(\ref{eq:forma}).

\medskip\noindent{\bf Case (c):}
For this case $l$ crosses the bucket at $x_{1,l}$ and $x_{2,l}$. The
integral over the shaded region above the line is given by:
\begin{equation}\label{eq:integrale}
    p_e = \int\limits_{x_{1,l}}^{\frac{x_{2,l}-x_{2,q}}{-t}+x_{1,q}}
    \int\limits_{-t(x_1 - x_{1,q}) + x_{2,q}}^{x_{2,u}}
    F_i~dx_2 dx_1 ~+~
    \int\limits_{\frac{x_{2,l}-x_{2,q}}{-t}+x_{1,q}}^{x_{1,l}}
    \int\limits_{x_{2,l}}^{x_{2,u}}
    F_i~dx_2 dx_1.
\end{equation}
The solution
has the form of Equation~(\ref{eq:forma}).

\medskip\noindent{\bf Case (d):}
For this case $l$ crosses the bucket at $x_{1,u}$ and $x_{2,l}$. The
integral over the shaded region is given by:
\begin{equation}\label{eq:integralf}
    p_f = \int\limits_{\frac{x_{2,l}-x_{2,q}}{-t}+x_{1,q}}^{x_{1,u}}
    \int\limits_{-t(x_1 - x_{1,q}) + x_{2,q}}^{x_{2,u}}
    F_i~dx_2 dx_1 ~+~
    \int\limits_{x_{1,l}}^{\frac{x_{2,l}-x_{2,q}}{-t}+x_{1,q}}
    \int\limits_{x_{2,l}}^{x_{2,u}}
    F_i~dx_2 dx_1.
\end{equation}
The solution
has the form of Equation~(\ref{eq:forma}).

\medskip\noindent{\bf Case (e):}
For this case $l$ crosses the bucket at $x_{1,l}$ and $x_{1,u}$. The
integral over the shaded region is given by:
\begin{equation}\label{eq:integralc}
    p_c = \int\limits_{x_{2,l}}^{x_{2,u}}
    \int\limits_{x_{1,l}}^{\frac{x_2 - x_{2,q}}{-t} + x_{1,q}}
    F_i~dx_1 dx_2.
\end{equation}
The solution
has the form of
\begin{equation}\label{eq:formc}
    c + \frac{d}{t} + \frac{e}{t^2}
\end{equation}
which is like Equation~(\ref{eq:forma}) with $a=b=0$.

\medskip\noindent{\bf Case (f):}
Similar to case(e), $l$ crosses the bucket at $x_{1,l}$ and
$x_{1,u}$. The integral over the shaded region is given by:
\begin{equation}\label{eq:integrald}
    p_d = \int\limits_{x_{2,l}}^{x_{2,u}}
    \int\limits_{\frac{x_2 - x_{2,q}}{-t} + x_{1,q}}^{x_{1,u}}
    F_i~dx_1 dx_2.
\end{equation}
The solution
has the form of Equation~(\ref{eq:formc}).

\medskip\noindent{\bf Case (g):}
For this case $l$ crosses the bucket at $x_{1,l}$ and $x_{1,u}$. The
integral over the shaded region is given by:
\begin{equation}\label{eq:integralg}
    p_g = \int\limits_{x_{1,l}}^{x_{1,u}}
    \int\limits_{-t(x_1 - x_{1,q})+x_{2,q}}^{x_{2,u}}
    F_i~dx_2 dx_1.
\end{equation}
The solution
has the form
\begin{equation} \label{eq:formg}
    at^2+bt+c
\end{equation}
which is like Equation~(\ref{eq:forma}) with $d=e=0$.

\medskip\noindent{\bf Case (h):}
The line $l$ crosses below all the corner vertices hence the
integral of the function is given as:
\begin{equation}\label{eq:integralh}
    p_h = \int\limits_{x_{1,l}}^{x_{1,u}}
    \int\limits_{x_{2,l}}^{x_{2,u}}
    F_i~dx_2 dx_1.
\end{equation}
The solution
has the form of Equation~(\ref{eq:formg}).


The above cases have solutions for each view in the form of
Equation~(\ref{eq:forma}). Hence the percentage function for a
single bucket as a function of $t$ is of the form:
\begin{align}\label{eq:BucketProbability}
    p &=\left( a_x t^2 + b_x t + c_x + \frac{d_x}{t} + \frac{e_x}{t^2} \right)
        \left( a_y t^2 + b_y t + c_y + \frac{d_y}{t} + \frac{e_y}{t^2} \right)\nonumber \\
      &~~~~\left( a_z t^2 + b_z t + c_z + \frac{d_z}{t} + \frac{e_z}{t^2} \right)
\end{align}
where $t\neq0$ when $d_x,d_y,d_z,e_x,e_y,e_z \neq 0$. Finally,
renaming variables gives the general form:
\begin{equation}\label{eq:BucketGeneralForm}
    p=a_6 t^6 + a_5 t^5 + a_4 t^4 + a_3 t^3 + a_2 t^2 + a_1 t + c +
    \frac{d_1}{t} + \frac{d_2}{t^2} + \frac{d_3}{t^3} + \frac{d_4}{t^4} +
    \frac{d_5}{t^5} + \frac{d_6}{t^6}
\end{equation}
where $t \neq 0$ when $d_i \neq 0$ for $1 \leq i \leq 6$. Since
Equation~(\ref{eq:BucketGeneralForm}) is closed under subtraction,
$\triangle p$ from Equation~(\ref{eq:percentofbucket2}) will also
have the same form.\medskip

As the {\em query space} from Definition~\ref{def:queryspace} sweeps
through a bucket, it crosses the bucket corner vertices. Each time a
corner vertex crosses the {\em query space} boundary, the case that
applies may change in one or more of the views.

\begin{definition}[Bucket and Index Time-Intervals]\label{def:buckettimeinterval}
The span of time in which no vertex from bucket $B_i$ enters or
leaves the query space defines a {\em bucket time-interval}. We
denote the time-interval as a half-open interval $[l,u)$ where $l$
is the lower bound and $u$ is the upper bound. Each {\em bucket
time-interval} has an associated percentage function $\triangle p$
given by Equation~(\ref{eq:percentofbucket2}). We define the {\em
index time-interval} similarly except that the span of time is
defined when no vertex from {\em any} bucket in the index enters or
leaves the query space.
\end{definition}

As we will see, index time-intervals are created from individual
bucket intervals. Throughout the rest of this dissertation we use
the term {\em time intervals} when the context clearly identifies
which type we mean.

\begin{definition}[Time-Partition Order]\label{def:timepartitionorder}%
Let $B$ be the set of buckets. Let $Q_1$ and $Q_2$ be two query
points and $(t^[,t^])$ be the query time interval. We define the
{\em Time-Partition Order} to be the set of ordered time instances
$TP={t_1,t_2,...,t_i,...,t_k}$ such that $t_1=t^[$ and $t_k=t^]$,
and each $[t_i,t_{i+1})$ is an {\em index time-interval}.
\end{definition}

\begin{example}[Calculating Bucket Time-Intervals]
    \label{ex:TimeIntervals} \rm %
    Continuing Example~\ref{ex:BuildIndex}, let $Q$ be a query defined
    by:
    \begin{eqnarray}
        q_{1} &=& (9.5,~8,~9.5,~8,~9.5,~8)\\
        q_{2} &=& (8.5,~5,~8.5,~5,~8.5,~5)\\
        T &=& (0.1,~10)
    \end{eqnarray}
    where $q_{1}$ and $q_{2}$ form the query space over the query time
    interval $T$. To determine time intervals when corner vertices do
    not change, find the slopes of lines through both query points and
    each corner vertex of the bucket. Figure \ref{fig:CornerLines} shows
    lines from the two query points to the corner vertices for the first
    dimension. Since the query points are the same in each dimension each
    will appear the same.
    \begin{figure}[t]
        \centering
        \includegraphics[width=3.75in]{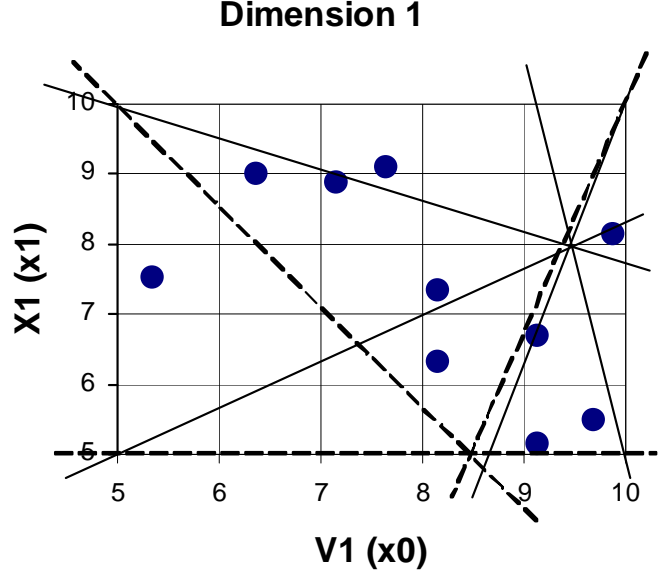}
        \caption{Lines from query points to corner vertices.}%
        \label{fig:CornerLines}
    \end{figure}
    The set of times when lines through $q_1$ (shown as solid lines) cross
    corner vertices is $\{0.\overline{4}, 6\}$. The set of times when
    lines through $q_2$ (shown as dotted lines) cross corner
    vertices and are in the time interval is $\{1.42857\}$. The
    union of these two sets along with the end points makes up the
    times used to create the time intervals:
    $\{(.1,0.\overline{4}),(0.\overline{4},1.42857),(1.42857,6),(6,10)\}$.
\end{example}
\medskip

Integration over the {\em spatial dimensions} of the eight possible
cases presented in Figure~\ref{fig:cases} gave a function of the
form of Equation~(\ref{eq:BucketGeneralForm}). {\em Maximizing}
Equation~(\ref{eq:BucketGeneralForm}) in the {\em temporal
dimension} by first taking the derivative, we get:
\begin{eqnarray}
    \triangle p'&=&(6a_{6}t^{12}+5a_{5}t^{11}+4a_{4}t^{10}+3a_{3}t^{9}+2a_{2}t^{8}+a_{1}t^{7} \nonumber \\
                &~&~ -d_{1}t^{5}-2d_{2}t^{4}-3d_{3}t^{3}-4d_{4}t^{2}-5d_{5}t -6d_{6}) / t^7 \label{eq:Derivative}
\end{eqnarray}
where $t \neq 0$. Solving $\triangle p'=0$ requires finding the
roots of this $12$-degree polynomial, {\em which is not possible
using an exact method}. Hence we need a numerical method for solving
the polynomial.

The following factors influenced the choice of the numerical method:
\begin{enumerate}
    \item Speed of the algorithm is more important than accuracy
          because we don't expect the original function to change
          dramatically over an index time-interval. We expect small
          change because in practice the time intervals are short.
    \item The algorithm must converge toward a solution within the
          interval, that is the algorithm must be stable.
    \item Given that we are maximizing Equation~(\ref{eq:BucketGeneralForm})
          over a short time interval, we don't expect
          Equation~(\ref{eq:Derivative}) to have more than one
          solution. This assumption may seem naive, but it is
          reasonable given factor (1).
\end{enumerate}

Factor (1) above is related to (3) in that it indicates that points
close together have similar values, but emphasizes that speed is the
goal. Factor (2) above eliminates several algorithms from
consideration, but must be required to keep from choosing a solution
that is not within the time interval evaluated.

Of the three points to consider, (3) is probably the least
intuitive. Consider the following conjecture:

\begin{conjecture}\label{lem:NearMaximums}
Given $p$ for a set of buckets, if the Euclidean distance between
two maxima is small, then the difference between the maxima is
small.
\end{conjecture}

Consider the physical characteristics of the system. The value of
$p$ over the time interval changes no more than $b_i$ for any bucket
$B_i$. Clearly $p$ either increases as it encompasses more of the
bucket or decreases at as it encompasses less of the bucket. When
$p$ represents the distribution over several buckets, each bucket
contributes a decreasing or increasing amount over the time
interval. Clearly $p$ is bounded below by $0$ and above by
$\sum\limits_i b_i$. Hence, the rate at which the derivative $p'$
changes is characterized by the physical system and reflects the
differences in the buckets as $t$ changes. Since $p$ does not change
dramatically over $t$ for any bucket, then change in several buckets
over $t$ will likewise not be dramatic. Hence if the distance
between two maxima is small, the maxima have a small difference in
magnitude. {\em This rational for the conjecture above is verified
by the experiments}.\medskip

Based on these factors, we use a common method for the first
approximation: we look at the graph of $p'$. Programmatically check
$c$ intervals of Equation~(\ref{eq:Derivative}) for a change in
sign. If there exists a sign change, use the bisection method to
find the root. If two points lie within $\epsilon$ of $0$, we
perform a check for each of these intervals when no change of sign
is found. If some roots exist, we check them for maximal values
along with the end points.

\begin{lemma}\label{lem:ConstRunningTimeForTimeInterval}
The approximate maximum within a time interval can be found in
$O(1)$ time.
\end{lemma}

\begin{proof}
Each {\em time interval} has an associated probability function
$\triangle p$ which is calculated in $O(1)$ time. Finding $\triangle
p' = 0$ also takes $O(1)$ time. By placing a constant bound on the
number of iterations in the bisection method, we bound the time
required in the numerical section of the algorithm by a constant.
Plugging in the solution found by the bisection method along with
the end points also takes $O(1)$ time. Hence, the running time to
find the maximum within a bucket is $O(1)$.
\end{proof}

We chose to limit the number of iterations in the bisection method
to 10, which limits the running time to a small constant value. This
value was chosen based on empirical observation that index
time-intervals remain small (about $0.01$ to $4$). Hence, using the
bisection method allows us to narrow our search down to an interval
at least as small as $\frac{1}{256}$ units of time. If time is
measured in hours, this interval equates to only $14$ seconds.

\begin{example}[Building Time-Intervals and Finding {\sc MaxCount}]\rm %
    Continuing Example~\ref{ex:TimeIntervals} we build the functions for
    time intervals
    \begin{equation}
        \{(.1,0.\overline{4}),(0.\overline{4},1.42857),(1.42857,6),(6,10)\}
    \end{equation}
    by integrating using the different cases from
    Figure~\ref{fig:cases}. For space concerns we omit the integrals here and
    note that the result of integrating each interval and finding
    the maximum gives a maximum of approximately $3$ at
    $t=0.\overline{4}$

    \noindent\textbf{Time Interval: }$[0.1,0.\overline{4}]$. Here case
    (c) holds for query point $q_2$ over this time interval. Hence the
    integral for query point $q_{2}$ and $t\in\lbrack.1,.\overline{4}]$
    in each dimension is given as:
    \begin{eqnarray}
        p_{c} &=& c\int_{8.5}^{10}\int_{5}^{10}2(0.7x_{0}-2.9)dx_{1}dx_{0}+\int_{5}^{8.5}\int_{-t(x_{0}-8.5)+5}^{10}2(0.7x_{0}-2.9)dx_{1}dx_{0}\nonumber\\
            &=& 117.5-17.354\bar{6}t \label{eq:Q2CaseEinterval1}
    \end{eqnarray}
    Case (g) holds for query point $q_1$ and thus the integral for query
    point $q_{1}$ and $t\in (.1,.\overline{4})$ in each dimension is
    given as:
    \begin{align}
        p_{g} &  =c\int_{5}^{10}\int_{-t(x_{0}-9.5)+8}^{10}2(0.7x_{0}-2.9)dx_{1}dx_{0}\nonumber\\
            &  =47.0-32.41\bar{6}t
    \end{align}
    Hence the integral of the region is:
    \begin{align}
        p &  =c\left(  p_{c}-p_{g}\right)  ^{3}\nonumber\\
        &  =2.106\times10^{-3}t^{3}+2.957\times10^{-2}t^{2}+0.138t+0.216
    \end{align}
    Evaluating $p$ at the start and end of the time interval we have
    $p(0.1)\approx0.23$ and $p(0.\overline{4})=0.28$. Figure
    \ref{fig:Interval1} shows $p$ in the time interval. Clearly $p$ is
    increasing and consequently we have a maximum at the end point
    $t=0.\overline{4}$.
    \begin{figure}[h]
        \centering
        \includegraphics[width=4in]{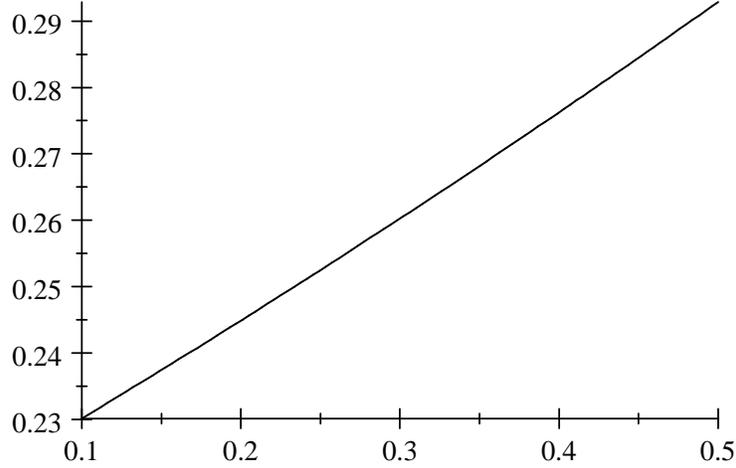}
        \caption{Graph of $p$, $0.1 \leq t \leq 0.\overline{4}$.}
        \label{fig:Interval1}
    \end{figure}
    Since there are $10$ points we must multiply
    $p(0.\overline{4})$ by $10$ to get the approximation for the time
    interval as:%
    \begin{equation}
        MaxCount_{0.1\leq t\leq 0.\overline{4}} \approx 2.8.
    \end{equation}
    Since we can not have partial points, we can round this result
    to $3$.\medskip

    The rest of the intervals are similar using different cases. We
    omit the remaining cases to save space and to eliminate the risk
    of boring the reader. None of the other intervals has a higher
    \sc{MaxCount} and so it follows that {\sc MaxCount} has an
    approximate value of $3$ at time $t=0.\overline{4}$.
\end{example}

\subsection{Dynamic {\sc MaxCount} Algorithm}\label{ssec:MaxCountAlgorithm}

\progstart \vspace{-18pt}
\begin{tabbing}
\hspace*{.5in}\=\hspace*{.5in}\=\hspace*{.5in}\= \kill
{\bf {\sc MaxCount}$(H, Q_1, Q_2, t^[,t^])$} \\
{\bf input:} \>\> A set of buckets $H$ built by the index structure presented, \\
             \>\> query points $Q_1(t)$ and $Q_2(t)$ and a query time interval $(t^[,t^])$. \\
{\bf output:}\>\> The estimated {\sc MaxCount} value.\\
\\
01.  \> $TimeIntervals \leftarrow \emptyset $                            \` $O(1)$ \\
02.  \> {\bf for} $i \leftarrow 0$ {\bf to} $|H|-1$                                  \` $O(B)$ \\
03.  \>   \> $CrossTimes \leftarrow $\textsc{ CalculateCrossTimes}$(Q_1,Q_2,t^[,t^],H_i)$     \` $O(1)$ \\
04.  \>   \> {\bf for} $j \leftarrow 1$ {\bf to} $|CrossTimes|-1$                    \` $O(1)$ \\
05.  \>   \>   \>\textsc{Union}$(TimeIntervals,TimeInterval(t_{j-1}, t_{j})$      \` $O(1)$ \\
06.  \>   \> {\bf end for} \\
07.  \> {\bf end for}\\
\\
08. \> $TimeIntervals = $\textsc{ BucketSort}$(TimeIntervals)$                       \` $O(B)$ \\
09. \> $IndexTimeIntervals = $\textsc{ Merge}$(TimeIntervals)$                       \` $O(B)$ \\
10. \> {\bf for each} $IndexTimeInterval \in IndexTimeIntervals$               \` $O(B)$ \\
11. \>  \> \textsc{calculate}$(MaxCount, MaxTime, IndexTimeInterval)$             \` $O(1)$ \\
12. \> {\bf end for} \\
\\
13. \> {\bf return} $(MaxCount, MaxTime)$
\end{tabbing}
\progend

The algorithm to compute {\sc MaxCount} with each line labeled with
its running time is given above. Line 01 initiates a set of bucket
time-interval objects to be empty. Line 03 returns a list of ordered
times when a line through $Q_1$ or $Q_2$ crosses a bucket corner
vertex. Line 05 turns this list into a set of $TimeInterval$ objects
and adds them to the set of $TimeIntervals$. We list this ``for
each'' loop as $O(1)$ because it consists of a constant number of
calculations bounded by the number of vertices in the bucket.  Line
08 uses the linear time sorting algorithm \textsc{BucketSort} to
sort the bucket time intervals. Line 09 creates the time-partition
order and index bucket time intervals from the bucket time intervals
in $O(B)$. An additional pass adds the bucket time intervals to the
appropriate index time-intervals in $O(B)$. Lines 10-12 perform the
{\sc MaxCount} calculation discussed above.

\medskip
In order to use the linear time \textsc{BucketSort} algorithm, we
need the following definition and lemmas.

\begin{definition}[Time-Interval Ordering]
\label{def:IntervalOrder}%
We define the lexicographical ordering $\prec$ of two {\em time
intervals} $A$ and $B$ as follows:
\begin{eqnarray}
    A.l < B.l                       & \Rightarrow & A \prec B \\
    A.l = B.l \quad \wedge \quad A.u < B.u & \Rightarrow & A \prec B \\
    A.l = B.l \quad \wedge \quad A.u = B.u & \Rightarrow & A = B
\end{eqnarray}

\end{definition}

\begin{figure}
  \centering
  \psfrag{Q}{{\tiny $Q$}}
  \psfrag{A1}{{\tiny $A=\frac{1}{2}$}}
  \psfrag{A2}{{\tiny $A=\frac{1}{4}$}}
  \psfrag{A3}{{\tiny $A=\frac{1}{12}$}}
  \includegraphics[height=3in]{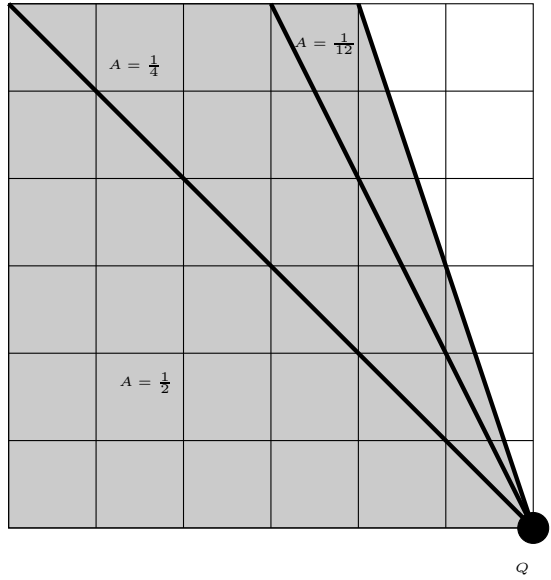}\\
  \caption{Areas of successive slopes.}
  \label{fig:SlopeDistribution}
\end{figure}

The distribution of time interval objects created in Line 08 of the
{\sc MaxCount} algorithm may not be uniform across the query time
interval $T=[t^[,t^]]$. However, we can still prove the following.

\begin{lemma}
\label{lem:TimeIntervalDistribution}%
If the distribution of buckets is uniform, then the distribution of
bucket time-interval objects can be uniformly distributed within the
sorting buckets of the bucket sort.
\end{lemma}
\begin{proof}
Consider the relationship between successive slopes measured as the
angles between lines through a query point $Q$ with slopes
$s_i=-t_i$ and $s_{i+1}=-t_{i+1}$. Suppose $\triangle t=1$ with
$t_0=0$ and $t_1=1$, then the angle between the two lines is
$\triangle s=\frac{\pi}{4}$. The solid lines in
Figure~\ref{fig:SlopeDistribution} show that half of the bucket
corner vertices are swept by the line sweeping through $Q$ between
$s_0=0$ and $s_1=-1$. Consider a query time interval $[0,10]$. Half
of the corner vertices, and thus half of the time intervals, are
between time $t=0$ and $t=1$. Thus, we conclude that the time
interval objects created by sweeping will not be uniformly
distributed throughout the query time interval.

Let $Q'$ be the midpoint between $Q_1$ and $Q_2$. Let $S =
\{t_1,...t_k\}$ where $t_1 = t^[$, $t_k=t^]$ and $t_{i+1} - t_i = L$
for some positive constant $L$ and $1 \leq i \leq k-1$. Let $D_B$ be
a bucket that contains the space in the 6-dimensional index. Model
the normalized bucket function for $D_B$ as a constant $F=1$. Thus
$p$, the bucket probability, from
Equation~(\ref{eq:BucketProbability}) becomes the hyper-volume of
the space swept by the line through $Q'$. By
Lemma~\ref{lem:ConstRunningTimeForTimeInterval}, we can find the
area for a specific time interval in $S$ in constant time. The
percentage of sorting buckets, $posb_i$, needed in any time interval
$T_i=[t_i,t_{i+1}] \in S$ within the query time interval is given
by:
\begin{equation}
    posb_i = \frac{p(t_{i+1})-p(t_i)}{p(t^])-p(t^[)}
\end{equation}
Let $N$ be the number of sorting buckets. Then, the number of
sorting buckets, $nosb_i$, assigned to interval $i$ is given by:
\begin{equation}
    nosb_i = N \cdot posb_i
\end{equation}
If $nosb_i<1$ we can combine it with $nosb_{i+1}$. If the query time
interval is very large, then we may need to include multiple time
intervals from $S$ to get one sorting bucket. Thus, we create more
sorting buckets (with smaller time intervals) in areas where the
expected number of bucket time intervals is large. Conversely, we
create fewer sorting buckets (with larger time intervals) in areas
where the expected number of bucket time intervals is small. Hence
we model each sorting bucket so that its time interval length
directly relates to the percentage of bucket time intervals that are
assigned to it. Thus, we conclude that we will uniformly distribute
the time interval objects across all sorting buckets.
\end{proof}

\begin{lemma}
\label{lem:BuscketSortConstantTimeInsertion}%
Insertion of any bucket time-interval object $T_O$ into the proper
sorting bucket can be done in $O(1)$ time.
\end{lemma}
\begin{proof}
The distribution of sorting buckets is determined by $k$ time
intervals in Lemma~\ref{lem:TimeIntervalDistribution}. Call these
{\em sorting time interval objects} where each object contains: the
lower bound $l$, the upper bound $u$, the number of sorting buckets
assigned to this interval $b_s$, the length of the time interval for
the sorting bucket $w$ and an array $B_p$ containing pointers to
these sorting buckets. Let $A$ be the array of sorting time interval
objects, and  $L$ be the length of each time interval where the time
intervals are as in Lemma~\ref{lem:TimeIntervalDistribution}. Then,
finding the correct sorting bucket for $T_O$ requires two
calculations:
\begin{eqnarray}
    SortingTimeInterval &=& A \left[ ~ \left\lfloor \frac{T_O.l}{L} \right\rfloor ~ \right] \\
    SortingBucket       &=& B_p \left[ ~ \left\lfloor \frac{T_O.l - SortingTimeInterval.l}{w} \right\rfloor ~ \right].
\end{eqnarray}
Each of these calculations requires constant time, hence $T_O$ can
be inserted into the proper sorting bucket in $O(1)$ time.
\end{proof}

Using the above two lemmas, we can prove the following.

\begin{theorem}
\label{th:constanttime} The running time of the {\sc MaxCount}
algorithm is $O(B)$ where $B$ is the number of buckets.
\end{theorem}

\begin{proof}
Let $H$ be the set of buckets where each bucket $B_i$ contains the
normalized trend function $F_i$. Let $Q_1$ and $Q_2$ be the query
points and $[t^[,t^]]$ be the query time interval. (Lines 01-07):
Calculating the time intervals takes $O(B)$ time because the cross
times for each bucket can be calculated in constant time. (Line 08):
By Lemmas~\ref{lem:TimeIntervalDistribution} and
\ref{lem:BuscketSortConstantTimeInsertion}, we have an approximately
even distribution of time interval objects within the sorting
buckets where we can insert an object in constant time. This result
fulfills the requirements of the \textsc{BucketSort},
\cite{IntroToAlgorithms}, which allows the intervals to be sorted in
$O(B)$ time. (Lines 09-12): Calculate the {\sc MaxCount} and time
for each time interval in constant time using
Lemma~\ref{lem:ConstRunningTimeForTimeInterval}. These lines takes
$O(B)$ time because there are $O(B)$ time intervals. Finding the
global {\sc MaxCount} and time requires retaining the maximum time
and count at line 11. Returning the {\sc MaxCount} and time takes
$O(1)$ time. Thus, the running time is given by $O(B) + O(B) + O(B)
+ O(1) = O(B)$.
\end{proof}

\subsection{An Exact {\sc MaxCount} Algorithm}\label{sec:ExactMaxCount}

The Exact MaxCount algorithm below finds the exact {\sc MaxCount}
values. It is easy to see that the running time is given by:
\begin{equation}\label{eq:ExactRunningTime}
    O(N) + O(n \log n)
\end{equation}
where $N$ is the number of points in the database and $n$ represents
the result size of the query.


It is possible to slightly improve the algorithm below. First,
divide the index space into $k$ subspaces and maintain separate
partial databases for each. Assign processes on individual systems
to each database to calculate the {\sc MaxCount} query and return
the time intervals to a central process. Merging the time interval
lists into a global time interval list saves time on the sorting
part of the algorithm. The running time for each of $k$ partial
databases would be close to $O(\frac{n}{k} \log \frac{n}{k})$. This
result is an approximate value because we do not guarantee an even
split between partial databases. Placing buckets for each partial
database in a \textsc{Tree} structure may be reasonable and could
cut down the average running time to $O(\log N + n \log n/k)$.
Implementation and analysis for this particular approach is left as
future work.

\progstart \vspace{-18pt}
\begin{tabbing}
\hspace*{.5in}\=\hspace*{.5in}\=\hspace*{.5in}\=\hspace*{.5in}\=
\kill
{\bf  {\sc ExactMaxCount}$(D, Q_1, Q_2, t^[, t^])$} \\
{\bf input:}  \>\> $D$ is the database of points. The query is made up of a \\
              \>\> hyper-rectangle $Q$ defined by points $Q_1$ and $Q_2$ and the time\\
              \>\> interval $T=[t^[, t^]]$ \\
{\bf output:} \>\> The exact {\sc MaxCount} and time at which it occurs. \\ \\
01.  \>$Times \leftarrow \emptyset$ //of \emph{CrossTime} objects      \` $O(1)$\\
02.  \>\textbf{for each} \emph{point} $p_i \in D$                      \` $O(N)$\\
03.  \>   \> \textbf{if} $p_i \in Q$ during $T$                        \` $O(1)$\\
04.  \>   \>   \> $EntryTime \leftarrow CalculateEntryTime(p_i,Q,T)$   \` $O(1)$\\
05.  \>   \>   \> $ExitTime \leftarrow CalculateExitTime(p_i,Q,T)$     \` $O(1)$\\
06.  \>   \>   \> \textbf{if} $EntryTime \in Times$                    \` $O(1)$\\
07.  \>   \>   \>   \> $Times.$\textsc{get}$(EntryTime).Count$++       \` $O(1)$\\
08.  \>   \>   \> \textbf{else} \\
09.  \>   \>   \>   \> $Times.$\textsc{add}$(new CrossTime(EntryTime))$\` $O(1)$\\
10.  \>   \>   \> \textbf{end if} \\
11.  \>   \>   \> \textbf{if} $ExitTime \in Times$                     \` $O(1)$\\
12.  \>   \>   \>   \> $Times.$\textsc{get}$(ExitTime).Count$-\,-      \` $O(1)$\\
13.  \>   \>   \> \textbf{else} \\
14.  \>   \>   \>   \> $Times.$\textsc{add}$(new CrossTime(ExitTime))$ \` $O(1)$\\
15.  \>   \>   \> \textbf{end if} \\
16.  \>\textbf{end for} \\
17.  \>\textsc{Sort}$(Times)$                                          \` $O(n \log n)$\\
18.  \>\textsc{traverse}$(Times,time,Max\textrm{-}Count)$ //tracking time\` $O(N)$\\
\> \> \> \> \qquad \qquad \qquad \quad //and {\sc MaxCount} \\
19.  \>\textbf{return} (time,{\sc MaxCount}) \` $O(1)$
\end{tabbing}
\progend

\section{Threshold Operators}\label{sec:ThresholdOperators}

\progstart \vspace{-18pt}
\begin{tabbing}
\hspace*{.35in}\=\hspace*{.3in}\=\hspace*{.3in}\=\hspace*{.3in}\=
\kill
{\bf  {\sc ThresholdRange}$(H, Q_1, Q_2, t^[, t^], M)$} \\
{\bf input:}  \>\> A set of buckets $H$ build by the index structure presented, \\
              \>\> query points $Q_1(t)$ and $Q_2(t)$, a query time interval $[t^[, t^]]$, \\
              \>\> and $M$ is the threshold value \\
{\bf output:} \>\> The estimated set of time intervals where $R$ contains more \\
              \>\> than $M$ points.\\
\\
01 - 08 are the same as the {\sc MaxCount} algorithm.\\
09.  \> $TimeIntervals \leftarrow \emptyset$                                        \`$O(1)$ \\
10.  \> \textbf{for each} $TimeInterval \in TimePartitionOrder$                     \`$O(B)$ \\
11.  \>     \> $CMaxCount \leftarrow \textsc{calculate}(\textsc{MaxCount}, MaxTime, TimeInterval)$\`$O(1)$ \\
12.  \>     \> \textbf{if} $CMaxCount > M$                                          \`$O(1)$ \\
13.  \>     \>  \> $TimeIntervals \leftarrow TimeIntervals \bigcup TimeInterval$    \`$O(1)$ \\
14.  \>     \>  \textbf{end if}                                                              \\
15.  \> \textbf{end for}                                                                     \\
16.  \> $\textsc{Merge}(TimeIntervals)$                                             \`$O(B)$ \\
17.  \> \textbf{return} $TimeIntervals$
\end{tabbing}
\progend

The {\sc ThresholdRange} algorithm shown above and described in
Definition~\ref{def:ThresholdRange} relates to {\sc MaxCount} in the
way we calculate the aggregation. We maintain a running count to
find time intervals that exceed the threshold value $M$. If we set
the threshold value near the {\sc MaxCount} value ($M \rightarrow$
{\sc MaxCount}), {\sc ThresholdRange} finds a small interval
containing the {\sc MaxCount}. We demonstrate this in the
experimental results,
Section~\ref{sec:ExperimentalResults}.\smallskip

The {\sc ThresholdRange} algorithm is the same as {\sc MaxCount} up
to Line 08, and then collects different information from each
$TimeInterval$ starting in Line 10. This leads to the following
Theorem.

\begin{theorem}
\label{th:ThresholdConstantTime}%
The estimated {\sc ThresholdRange} query runs in $O(B)$ time.
\end{theorem}
\begin{proof}
The {\sc ThresholdRange} algorithm differs from the {\sc MaxCount}
algorithm only in lines 09-17. Lines 11-14 run in $O(1)$ time. Line
10 executes lines 11-13 $O(B)$ times. In line 16,
$\textsc{Merge}(TimeIntervals)$ is a linear walk of the time
intervals that joins adjacent time intervals $T_a$ and $T_b$ when
$T_a \bigcup T_b$ would form a continuous time interval. The
calculation is trivially $O(1)$ time for joining the adjacent
intervals. Hence, we conclude by Theorem~\ref{th:constanttime} that
the {\sc ThresholdRange} runs in $O(B)$ time.
\end{proof}

\subsection{Threshold: Sum, Count and Average}

We give the following three operators based on {\sc ThresholdRange}
and conclude that none of the changes to the algorithm affect the
running time of the {\sc ThresholdRange} algorithm.\medskip

\noindent {\sc ThresholdCount}: \\
By adding a line between 14 and 15 in the {\sc ThresholdRange}
algorithm that counts the merged time intervals, we can return the
count of time intervals during the query time interval where
congestion occurs. This count of time intervals gives a measure of
variation in congestion. That is, if we have lots of time intervals,
we expect that we have a large number of pockets of congestion.
Since {\sc ThresholdCount} does not give information relative to the
entire time interval, it may need to be examined in light of the
total time above the threshold.\medskip

\noindent {\sc ThresholdSum}: \\
By summing the times instead of using the $\bigcup$ operator in line
13 of the {\sc ThresholdRange} algorithm, we can return the total
congestion time during the query time interval. This total gives a
measure of the severity of congestion that may be compared to the
length of query time.\medskip

\noindent {\sc ThresholdAverage}: \\
By adding a line between lines 14 and 15 in the {\sc ThresholdRange}
algorithm that finds average length of the merged time intervals, we
can return the average length of time each congestion will last.
This average gives a different measure of the severity of each
congestion.\medskip



\subsection{Count Range Algorithm}

The {\sc CountRange} algorithm is an adaptation of {\sc MaxCount} in
that it is the {\sc Count} portion of the {\sc MaxCount} query.
Using the equations for the cases described in
Figure~\ref{fig:cases}, we calculate the {\sc CountRange} as
follows:


\begin{figure}[h]
\begin{minipage}{0.49\textwidth}
    \centering
    \psfrag{Q1}{$Q_1$}
    \psfrag{Q2}{$Q_2$}
    \psfrag{lq2t1}{$l_{Q_2,t^[}$}
    \psfrag{lq2t2}{$l_{Q_2,t^]}$}
    \psfrag{lq1t1}{$l_{Q_1,t^[}$}
    \psfrag{q1t2}{$l_{Q_1,t^]}$}
    \psfrag{x0lvxl}{$(x_{0,l},v_{x,l})$}
    \psfrag{x0uvxu}{$(x_{0,u},v_{x,u})$}
    \includegraphics[width=.8\textwidth]{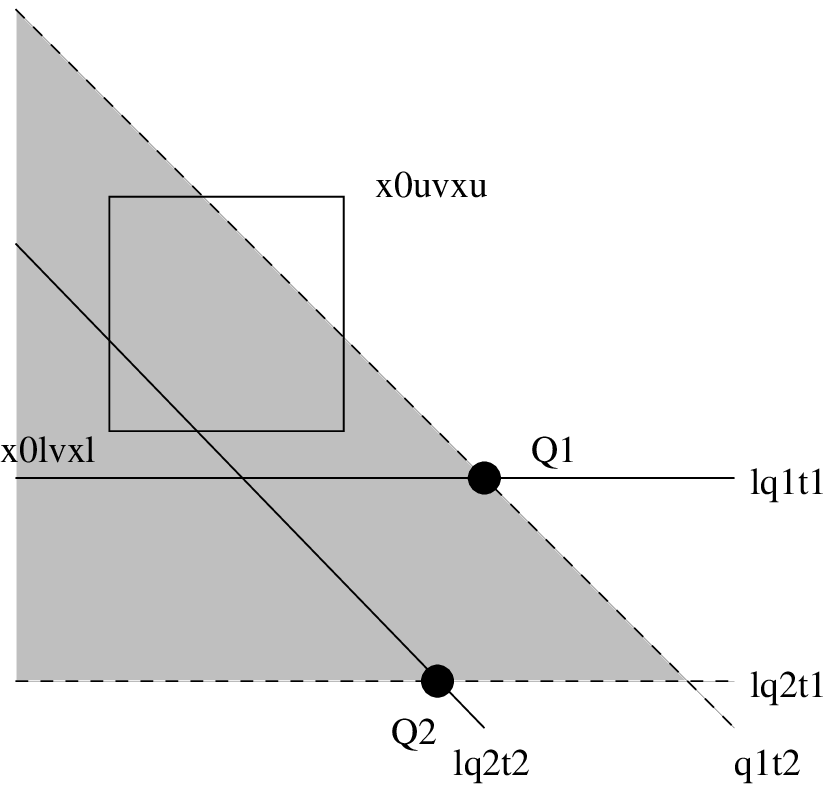}\\
    \caption{{\sc CountRange} $Q_1$ at $t^{]}$ to $Q_2$ at $t^{[}$.}
    \label{fig:CountRangeNormal1}
\end{minipage}
\begin{minipage}{0.49\textwidth}
    \noindent
    \psfrag{Q1}{$Q_1$}
    \psfrag{Q2}{$Q_2$}
    \psfrag{lq2t1}{$l_{Q_2,t^[}$}
    \psfrag{lq2t2}{$l_{Q_2,t^]}$}
    \psfrag{lq1t1}{$l_{Q_1,t^[}$}
    \psfrag{lq1t2}{$l_{Q_1,t^]}$}
    \psfrag{x0lvxl}{$(x_{0,l},v_{x,l})$}
    \psfrag{x0uvxu}{$(x_{0,u},v_{x,u})$}
    \includegraphics[width=.8\textwidth]{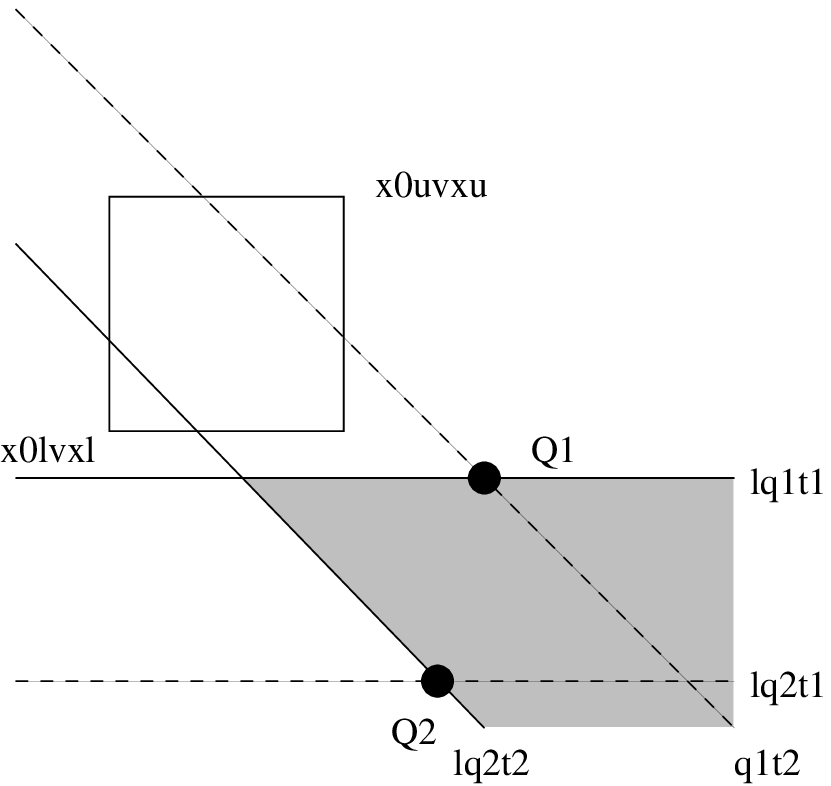}\\
    \caption{{\sc CountRange} $Q_1$ at $t^{[}$ to $Q_2$ at $t^{]}$.}
    \label{fig:countRangeNormal2}
\end{minipage}
\end{figure}

For each bucket we determine if the bucket is completely in or
completely out of the query space. First we find the beginning and
ending time intervals. For each time interval, we get the associated
function $\triangle p$ given in Equation~(\ref{eq:percentofbucket2})
and its components. The components $\triangle p$ given in
Equation~(\ref{eq:percentofbucket}) define the area above a line
through $Q_1$ and $Q_2$ at times $t^[$ and $t^]$.
Figures~\ref{fig:CountRangeNormal1} and \ref{fig:countRangeNormal2}
show these four lines. Figure~\ref{fig:CountRangeNormal1} shows the
shaded area defined by:
\begin{equation}\label{eq:pleft}
    \triangle \overleftarrow{p} = p_{Q_2,t^[} - p_{Q_1,t^]}.
\end{equation}
Figure~\ref{fig:countRangeNormal2} shows the shaded area:
\begin{equation}\label{eq:pright}
    \triangle \overrightarrow{p} = p_{Q_2,t^]} - p_{Q_1,t^[}.
\end{equation}
If $\triangle \overleftarrow{p}$ or $\triangle \overrightarrow{p}$
for bucket $i$ is equal to the count of the bucket, then bucket $i$
is completely contained in the query. If $\triangle
\overleftarrow{p}$ and $\triangle \overrightarrow{p}$ for bucket $i$
are equal to $0$, then bucket $i$ is not contained in the query. If
neither of these is true, we approximate the count for bucket $i$ as
the $\max (\triangle \overleftarrow{p}, \triangle
\overrightarrow{p})$. That is, we calculate the number of points in
bucket $i$ that contribute to the {\sc CountRange} as:
\begin{equation}\label{eq:CountRangei}
    count_i = \left\{
    \begin{array}{llr}
        b_i & \textrm{ if } & \triangle \overleftarrow{p} = b_i \vee \triangle \overrightarrow{p} = b_i \\
        0   & \textrm{ if } & \triangle \overleftarrow{p}=\triangle \overrightarrow{p} = 0 \\
        \max (\triangle \overleftarrow{p}, \triangle \overrightarrow{p}) & & \textrm{ Otherwise}
    \end{array}
    \right.
\end{equation}
This calculation requires that we keep the single dimension
equations for $Q_1$ and $Q_2$ available and not discard them after
finding $\triangle p$ (see Equation~(\ref{eq:percentofbucket2})).

Hence, we have the following algorithm for {\sc CountRange}:

\progstart\vspace{-18pt}
\begin{tabbing}
\hspace*{.5in}\=\hspace*{.5in}\=\hspace*{.5in}\=\hspace*{.5in}\=
\kill
{\bf  {\sc CountRange}$(H, Q_1, Q_2, t^[,t^])$} \\
{\bf input:}  \>\> A set of buckets $H$ built by the index structure presented, \\
              \>\> query points $Q_1(t)$ and $Q_2(t)$ and a query time interval $(t^[,t^])$. \\
{\bf output:} \>\>the estimated {\sc CountRange}. \\
\\
1.\>  $Count \leftarrow 0$                                             \` $O(1)$ \\
2.\>  \textbf{for each} \emph{bucket} $B_i \in D$                                      \` $O(B)$ \\
3.\>  \>  \textsc{Calculate}($\triangle \overleftarrow{p}, \triangle \overrightarrow{p}$) //using Equations~(\ref{eq:pleft})-(\ref{eq:pright}) \` $O(1)$ \\
4.\>  \>  \textsc{Calculate}($count_i$) //using Equation~(\ref{eq:CountRangei})   \` $O(1)$ \\
5.\>  \>  $Count \leftarrow Count + count_i$                                           \` $O(1)$ \\
6.\>  \textbf{end for} \\
7.\>  \textbf{return} $Count$ \` $O(1)$
\end{tabbing}
\progend

\begin{theorem}
\label{th:RangeConstantTime} The {\sc CountRange} query runs in
$O(B)$ time.
\end{theorem}
\begin{proof}
Consider two different data structures for our buckets:
\textsc{HashTables} and \textsc{R-trees}. In the case of indexing
using an \textsc{R-tree}, the worst case requires that we examine
all buckets used in generating {\sc CountRange}. It is possible that
this list could include all $B$ buckets giving a worst case of
$O(B)$. In the case of using a \textsc{HashTable}, we must examine
all $B$ buckets. By Lemma \ref{lem:ConstRunningTimeForBucket}, and
because Equations~(\ref{eq:BucketGeneralForm}) and
(\ref{eq:CountRangei}) are calculated in constant time, each bucket
can be examined to determine the count that contributes to the {\sc
CountRange} query in constant time. Therefore, the algorithm runs in
$O(B)$ time.
\end{proof}

We note that {\sc CountRange} is a simplification of the {\sc
MaxCount} operator in that we do not examine every time interval.
Further we have a slightly different form of $\triangle p$ from
Equation~(\ref{eq:percentofbucket2}) to find the count.

\section{Experimental Results}\label{sec:ExperimentalResults}

We collected data from over $7500$ queries that were selected from a
set of randomly generated queries. The selection process weeded out
most similar queries and kept a set that represents narrow queries,
wide queries, near corner or edge queries, and queries outside the
space contained in the database. Throughout our experiments, we did
not see significant accuracy fluctuation due to any of these types
of queries.

Each experimental run consists of running all of the queries at
several different decreasing bucket sizes on a single data set. We
made experimental runs against data sets ranging from 10,000 points
to 1,500,000 points\footnote{Threshold aggregation runs go only to 1
million points at which we already achieve acceptable error.}.

In the following experimental analysis, we measure the percentage
error of the estimation algorithm relative to the exact-count
algorithm as follows:
\begin{equation}\label{eq:RelativeError}
    Error_{Relative} = \frac{|Exact~Operator-Estimated~Operator|}{Exact~Operator}
\end{equation}
Equation (\ref{eq:RelativeError}) provides a useful measure if the
query returns a reasonable number of points. Queries that return a
small number of points indicate that we should use the exact method.

For {\sc ThresholdRange}, we measure the percentage of intervals
given by the accurate algorithm not covered by the estimation
algorithm using the operator {\sc UC} for uncovered. That is, {\sc
UC}$(a,b)$ returns the sum of the lengths of intervals in $a$ not
covered by intervals in $b$. We divide the result by the accurate
{\sc ThresholdSum} to determine the {\sc ThresholdRange error}:
\begin{equation}\label{eq:ThresholdRangeError}
    \texttt{error} =
    \frac{\textsc{UC}\left(\textit{Ext. }\textsc{ThresholdRange}, \textit{Est. }\textsc{ThresholdRange}\right)}{\textit{Ext. }\textsc{ThresholdSum}}
\end{equation}
We also measure the percentage of intervals given by the estimate
algorithm not covered by the exact algorithm. We divide the result
by the estimated {\sc ThresholdSum} to determine the {\sc
ThresholdRange excess-error}.
\begin{equation}\label{eq:ThresholdRangeExcessError}
    \texttt{excess-error} = \frac{\textsc{UC}\left(\textit{Est. }\textsc{ThresholdRange} \backslash \textit{Ext. }\textsc{ThresholdRange}\right)}{\textit{Est. }\textsc{ThresholdSum}}
\end{equation}

We performed all the data runs on a Athlon 2000  with 1 GB of RAM.
During each of the queries the program does not contact the server
tier and, thus, minimizes the impact of running a server on the same
computer. The program pre-loads all data into data structures so
that even the exact algorithms do not contact the server tier.

\subsection{Data Generation}

\begin{figure}[h]
\centering
\begin{minipage}{6in}
\centerline{ \hspace{-1em}
\mbox{\includegraphics[width=2in]{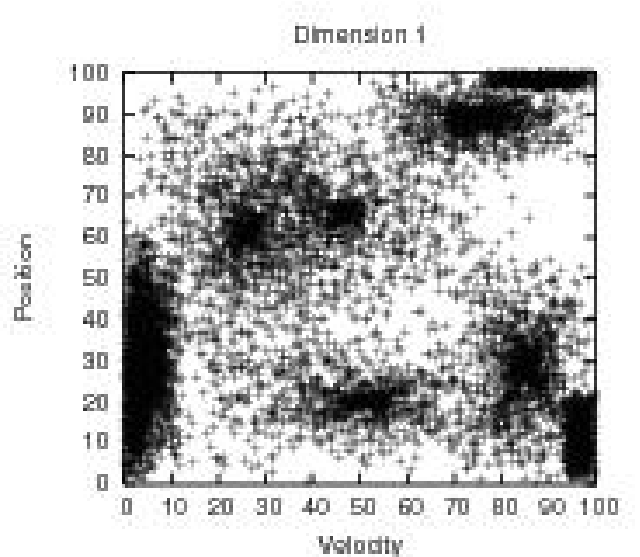}}\hspace{-0.1in}
\mbox{\includegraphics[width=2in]{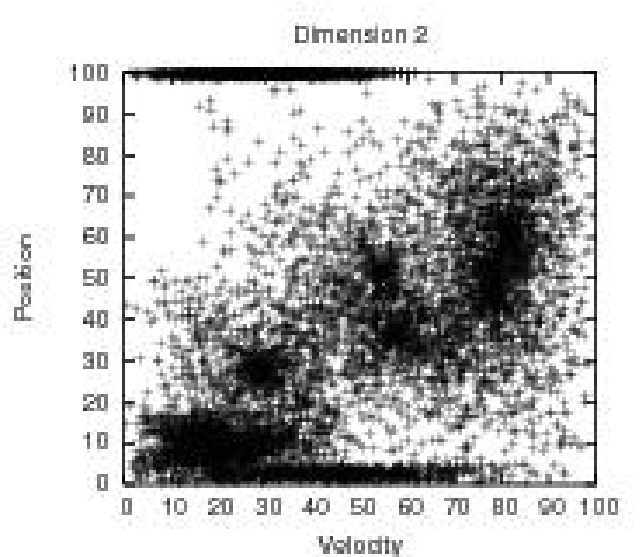}}\hspace{-0.1in}
\mbox{\includegraphics[width=2in]{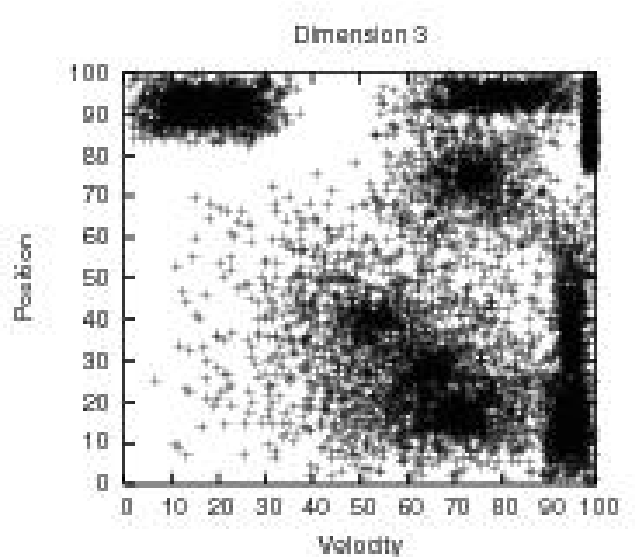}}}
\end{minipage}
\caption{$X$-View, $Y$-view and $Z$-view of sample data.}
\label{fig:sampledata}
\end{figure}

Data for the experiments was randomly generated around several
cluster centers. The $i^{th}$ point generated for the database is
located near a randomly selected cluster at a distance between $0$
and $d$, where $d$ is proportional to $i$. This method is similar to
the Ziggurat~\citep{marsaglia2000zmg} method of generating gaussian
(or normal) distributions used in the
GSTD~\citep{theodoridis1999gsd} and
G-TERD~\citep{tzouramanis2002gte} spatiotemporal data
generators~\citep{nascimento2003sar}. However, our method does not
generate strictly Gaussian distributions since the distributions may
stretch and compress along an axis. Our goal was to generate a
cluster that represents a source location and velocity that has most
elements starting near a center point and decreasing as one moves to
a boundary for the cluster. This method models source regions where
the objects all head about the same direction. A secondary goal was
to make certain that clusters were random in size and shape. The
program is also capable of approximating a Zipf distribution used in
\citep{CC02,Revesz20031,TSP03}. However, a single Zipf distribution
does not test the adaptability of our algorithm well. I.e. our
algorithm is capable of modeling a Zipf distribution and as such we
could use a single bucket. Figure~\ref{fig:sampledata} shows a
sample of a data set with points projected onto the three views. The
clusters look even more random, because they can overlay one
another. When one looks at these, they nearly resemble the lights of
a city from the air.

Along with a single Zipf distribution, we also note that a randomly
generated uniform-distribution is not a good distribution to use for
these types of experiments. Uniform distributions do not test the
ability of the algorithm to adapt. In fact from earlier experiments
in~\citep{Anderson20061} we have found that using such a
distribution gives great (though meaningless) results. The problem
resolves to a system capable (and willing to) model a uniform
distribution finding a nearly perfect uniform distribution to model.
Hence these results are neither realistic, nor meaningful.

\subsection{Parameter Effects}

The index space ranges from $0$ to $100$ in each dimension. The {\bf
number of points} in the different data sets ranges from $10,000$ to
$1,500,000$. The following parameters were used in creating the
index and finding the {\sc MaxCount}. \medskip

\noindent{\bf Size of Buckets:} The size of the buckets determines
the number of possible buckets in the index. In the experiments,
buckets divide the space up such that there are $5$ to $20$
divisions in each dimension\footnote{Some {\sc MaxCount} runs
included up to 40 divisions increasing accuracy, but not enough to
warrant the extra running time.}. These divisions equate to bucket
sizes ranging from $5$ to $20$ units wide in each dimension.
Relative to our previous work \citep{Anderson20061}, this algorithm
puts much more space into each bucket creating bigger buckets.
\medskip

\noindent{\bf Query Location:} Locating the query near the lower or
upper corners affects relative accuracy because the query returns
very few points. Queries in this region are not interesting because
they rarely involve many points and represent a query region that
moves away from points in the database or barely moves at all. The
small number of points returned indicates use of the exact
algorithms.
\medskip

\noindent{\bf Query Types:} In~\citep{Anderson20061}, we considered
queries with several different characteristics: dense, sparse, and
Euclidean distance as it related to bucket size. By modeling the
skew in buckets, we minimize the effect of these characteristics to
the point that they did not impact the query error. Queries where
the distance between the query points was small appeared to do as
well as wider queries {\em providing they returned a reasonable
number of points}. This result is a clear improvement over previous
work that assumed uniform density within a bucket.\medskip

\noindent{\bf Cluster Points:} Index space saturation determines the
number of buckets necessary for the index. The number of cluster
points does not appear to affect error as much as the space
saturation. Further, we do not consider a larger number of cluster
points reasonable since the index space approaches a uniform
distribution as the number of cluster points increases. Gaps
introduce difficult areas to model when they are not uniform. And
once again we reiterate, uniform distributions are not useful. In
our experiments cluster points number between 10 and 50. \medskip

\noindent{\bf Histogram Divisions:} Increasing histogram divisions
to $s>5$ had no affect on the accuracy. This result is not
unexpected because histograms are used to define a trend function
relative to trend functions on other axes. Increasing the histogram
divisions has a tendency to flatten the lines. However,
normalization flattens the trend function while maintaining the
relationships between trends and hence this behavior is easily
explained. Thus, increasing histogram divisions only increases the
running time without increasing accuracy.\medskip

\noindent{\bf Threshold Value:} The threshold value determines the
accuracy when set to low values compared to the number of points in
the database. As expected, these extreme point values produce
accurate estimations. High values also follow this trend.\medskip

\noindent{\bf Time Endpoints:} When dealing with either small time
end points or small buckets, the method is susceptible to rounding
error. In particular, Equation~(\ref{eq:BucketGeneralForm}) contains
both $t^6$ and $\frac{1}{t^6}$ terms. For very small values, on the
order of $1 \times 10^{-54}$ for 64-bit doubles, these calculations
are extremely sensitive and care must be given to guard against
rounding error. Those errors showed in two ways. First, by a direct
warning programmed into the solution, and second, by a series of
fairly stable time values for the {\sc MaxCount} followed by
unstable variations when increasing the number of buckets. At some
point, smaller bucket sizes increase the likelihood of errors in
both time and count values. Also smaller buckets contain fewer
points, which impacts the size of the constants in
Equation~(\ref{eq:BucketGeneralForm}). Hence, as the bucket size
becomes smaller in successive runs, the existence of instability in
the time values after a series of stable values predicts that an
accurate {\sc MaxCount} may be found in the previous larger bucket
size. {\em Throughout our experiments, this condition was an
excellent predictor of an accurate {\sc MaxCount}}.\medskip

\begin{figure}[htb]
  \centering
  \includegraphics[width=4.5in]{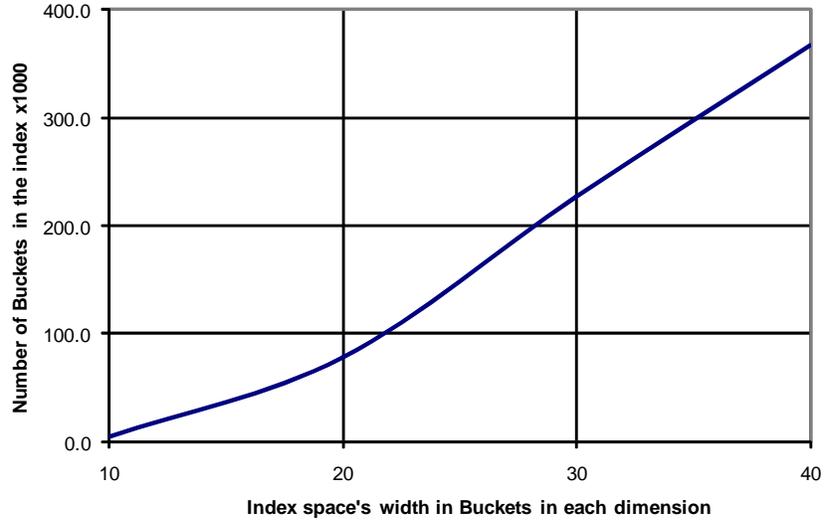}\\
  \caption{Ratio of the number of buckets in the index to the width of the space measured in buckets.}
  \label{fig:BucketsToPointsRatio}
\end{figure}

The experiments demonstrated that 6-dimensional space compounds the
problem when creating small buckets. Creating an index with unit
buckets would result in the possibility of having $1\times 10^{12}$
buckets. Clearly this number is unrealistic for common moving object
applications where we may be dealing with million(s) of objects. In
practice the number of buckets needed to reach acceptable error
levels was between $78,000$ and $227,000$ buckets. These numbers
reflect the ability to reach error levels under $5\%$ and were
roughly related to the saturation of the space by the points. It
should be clear that a higher saturation of the space by points
would require a larger number of buckets.
Figure~\ref{fig:BucketsToPointsRatio} shows that we had a roughly
linear increase in the number of buckets for an exponential increase
in the space. This pleasant surprise indicates that for unsaturated
data sets, the exponential explosion of space is manageable.

\subsection{Running Time Observations}

\begin{figure}[h]
  \centering
  \includegraphics[width=4.5in]{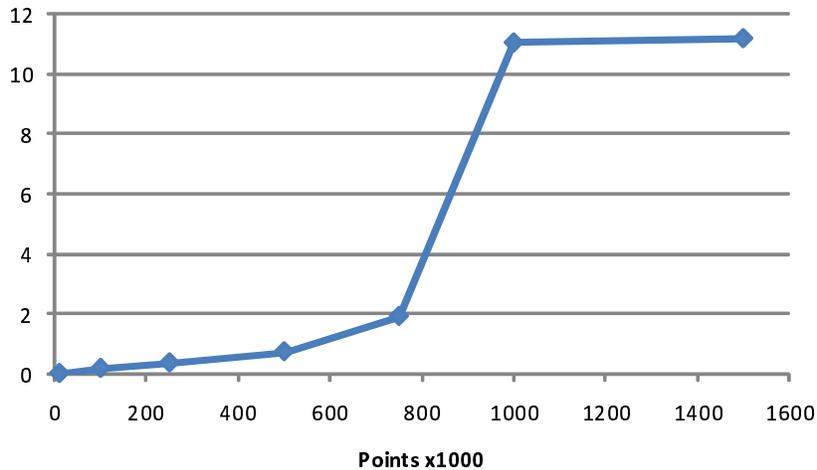}\\
  \caption{Ratio of exact running time to estimated running time.}
  \label{fig:runningtime}
\end{figure}

Figure~\ref{fig:runningtime} shows the average ratio of the exact
{\sc MaxCount} running time to the estimated {\sc MaxCount} running
time as a function of the number of points in the database. This
result shows a nearly exponential growth when comparing the values
between 10,000 and 1,000,000. The leveling off occurs because the
number of points returned by the queries of 1 million points nearly
equals the number of points returned by the queries of 1.5 million
points. This result precisely matches our running-time analysis of
the exact and estimation algorithms.

A natural question is when to use the exact versus the estimated
methods. In runs with a small number of points that need to be
processed, 
the exact and estimation methods run about equally fast. However,
when the result size reaches values greater than $40,000$ (our
experiments returned sets as large as 331,491), the estimation
algorithms run up to $35$ times faster than the exact algorithms.
Further, we note that the error is less predictable at smaller
results sizes. Hence for small databases or in queries that return
small result sets, efficiency and accuracy both indicate using the
exact method. However, for large data sets greater than or equal to
1 million points, the estimation method greatly out-performs the
exact method.

\subsection{Operator Observations}

As expected, we noticed that each operator runs in about the same
time as {\sc MaxCount}. Only error values seemed to be different
when studying different types of aggregation (e.g., when studying
overlap error in {\sc ThresholdRange} versus count error in {\sc
MaxCount}). Never-the-less, we have similarities between the
results. Almost all the figures in this section look like a view of
mountains from a valley. That is what we expected to see and the
lower and flatter the terrain the better. Buckets increase from back
to front and point set sizes increase from left to right.

\subsection{\sc MaxCount}

\begin{figure}[h]
  \centering
  \includegraphics[width=5.5in]{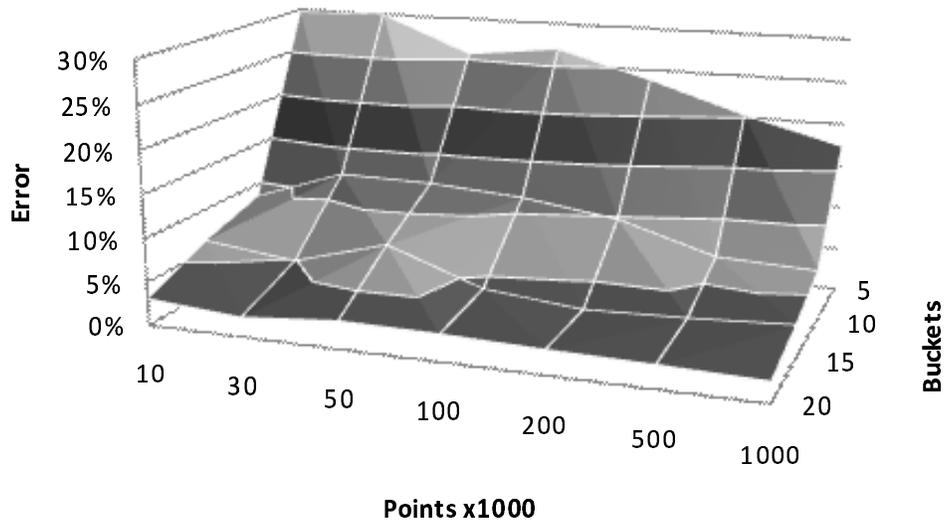}\\
  \caption{{\sc MaxCount} error.}
  \label{fig:RelativeError}
\end{figure}

Figure~\ref{fig:RelativeError} shows that increasing the number of
buckets to the indicated values dramatically decreases the {\sc
MaxCount} error. As the number of points increases we also see a
decrease in the error. Note that for larger buckets (e.g. smaller
values on the ``Buckets per Dimension axis''), the error decreases
at a slightly faster rate.

The exact {\sc MaxCount} provided the values against which our
estimation algorithm was tested for accuracy. Since the method does
not rely on buckets, and has zero error, we note only that on
queries with small result sizes, this method performs as well, or
better than the estimation algorithm.

\subsection{\sc ThresholdRange}

\begin{figure}[h]
  \centering
  \includegraphics[width=4in]{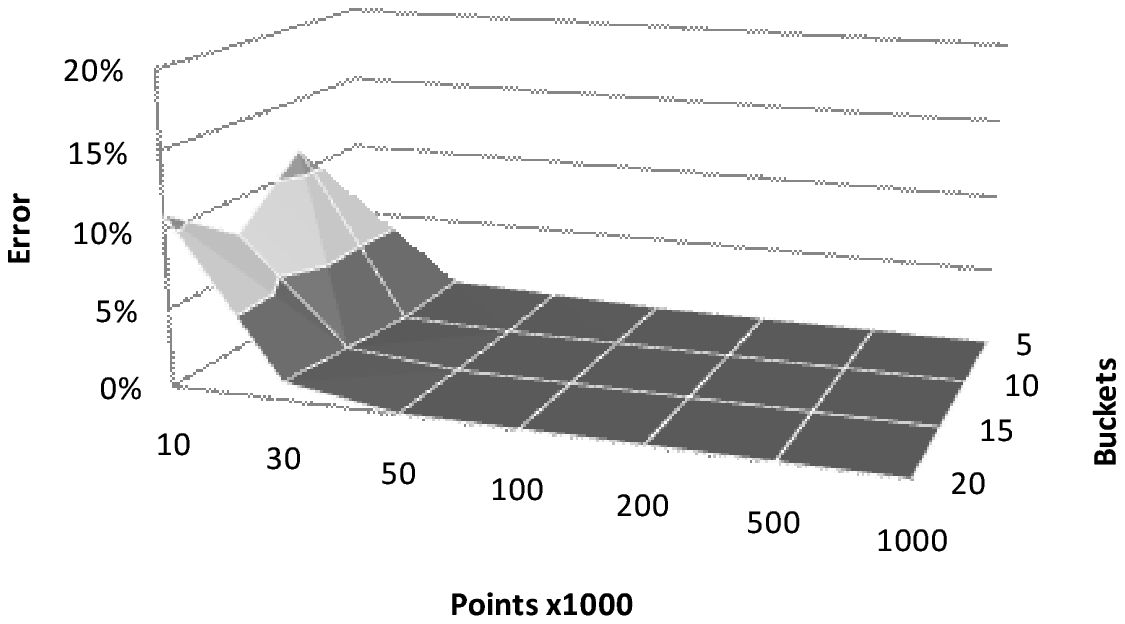}\\
  \caption{{\sc ThresholdRange} error.}
  \label{fig:TRE10}
\end{figure}

\begin{figure}[h]
  \centering
  \includegraphics[width=4in]{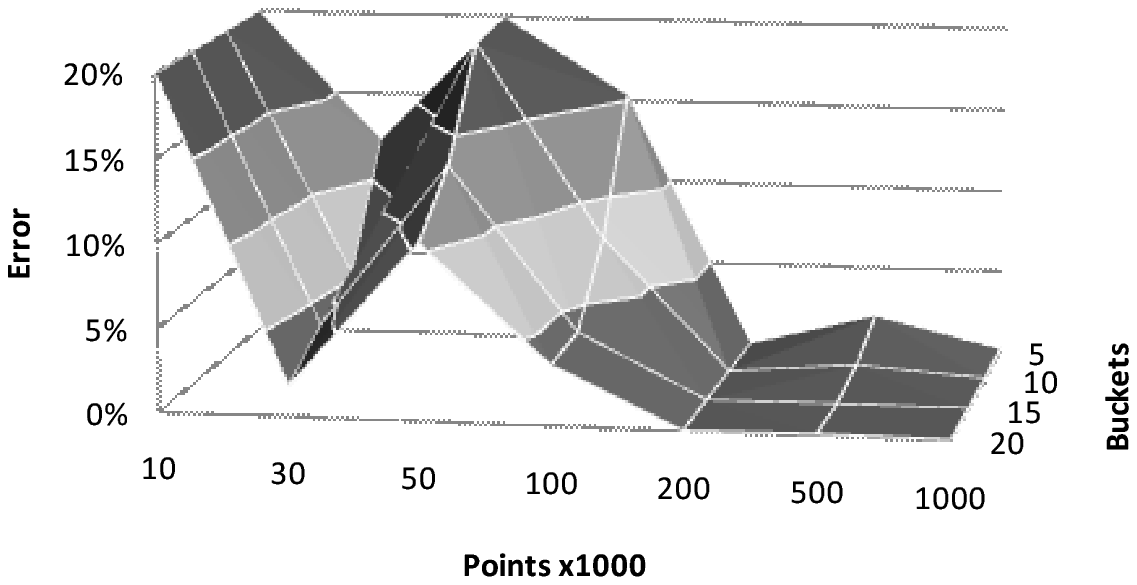}\\
  \caption{{\sc ThresholdRange} error.}
  \label{fig:TREE10}
\end{figure}

Figures~\ref{fig:TRE10} and \ref{fig:TREE10} give the {\sc
ThresholdRange} error and {\sc ThresholdRange} excess error
respectively for $T=10$. {\sc ThresholdRange} error gives the
percentage of the exact intervals not covered by the estimation
value, and {\sc ThresholdRange} excess error gives the percentage of
the estimation not covering the exact. These figures show that our
method acts conservatively in covering more than is needed. However,
at larger point-set sizes, we still achieve under 5\% error.
Figure~\ref{fig:TRE10} shows 0\% error caused by the point count
staying above 10\% in data sets containing more than 30,000 points.
Figure~\ref{fig:TREE10} shows that we covered at least 10\% more
time in the query time interval than needed until we reach larger
point sets. Still, we showed improvement with more buckets.

\begin{figure}
  \centering
  \includegraphics[width=4in]{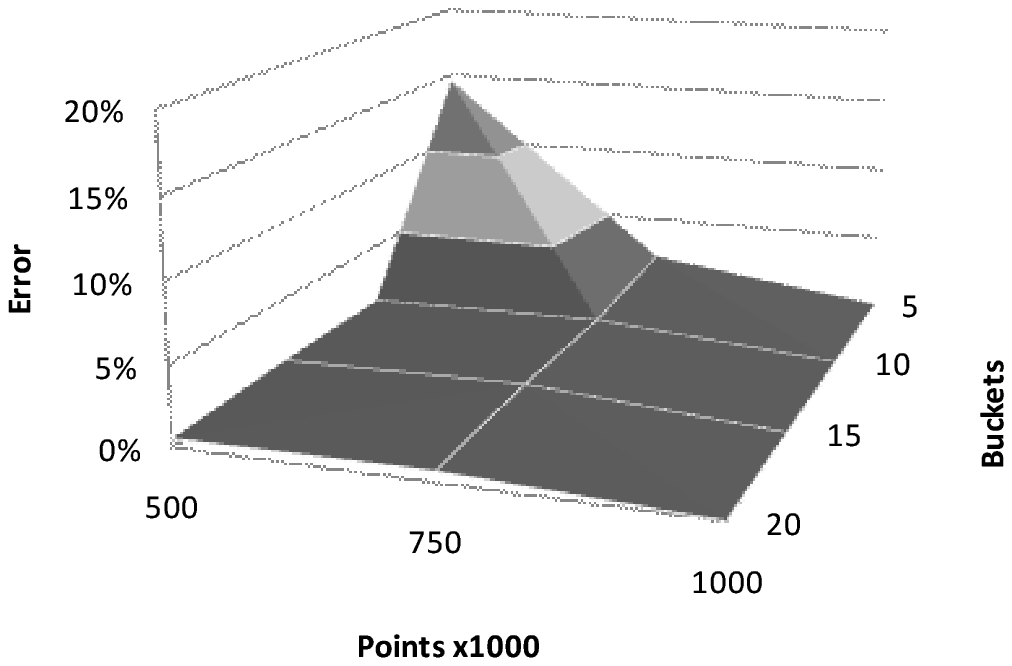}\\
  \caption{{\sc ThresholdRange} error, T=1000.}
  \label{fig:TRE1000}
\end{figure}

\begin{figure}
  \centering
  \includegraphics[width=4in]{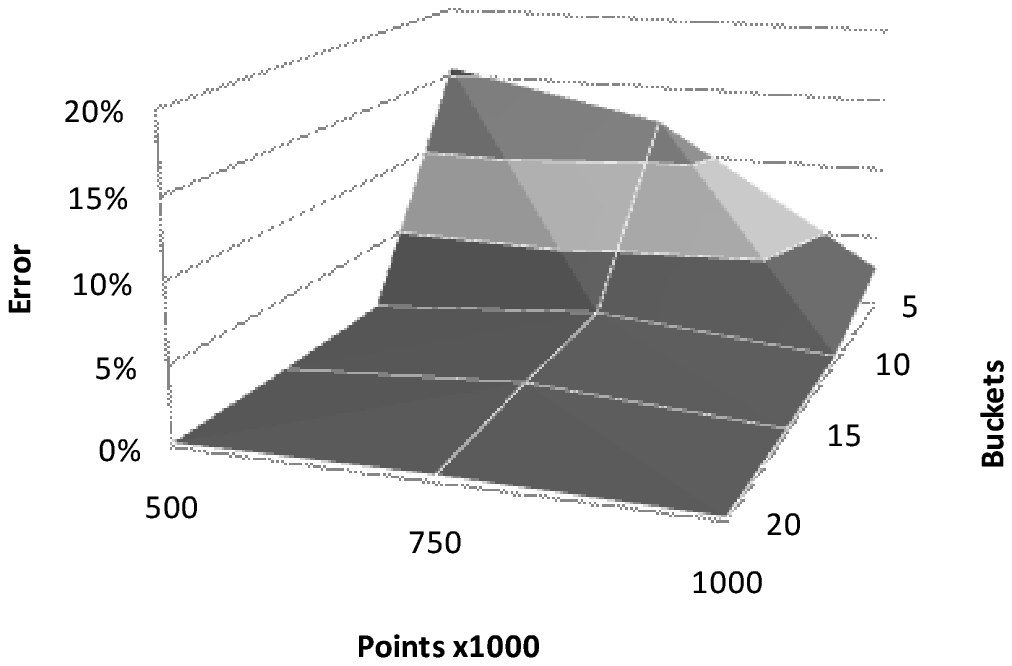}\\
  \caption{{\sc ThresholdRange} excess error, T=1000.}
  \label{fig:TREE1000}
\end{figure}

At $T=1000$, we see 0\% error until we reach point sets of 500,000
and greater. Figure~\ref{fig:TRE1000} shows excellent results with
buckets above 10. Also, Figure~\ref{fig:TREE1000} shows that the
excess error drops to near 0\% as well.

\begin{figure}
  \centering
  \includegraphics[width=4in]{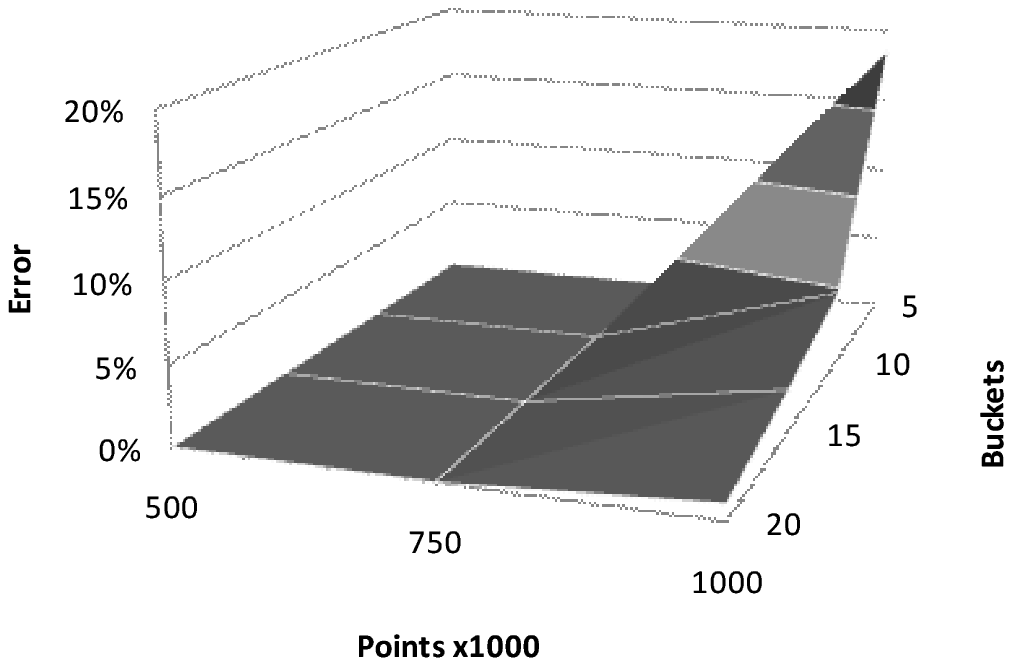}\\
  \caption{{\sc ThresholdRange} error, T=100000.}
  \label{fig:TRE100000}
\end{figure}

\begin{figure}
  \centering
  \includegraphics[width=4in]{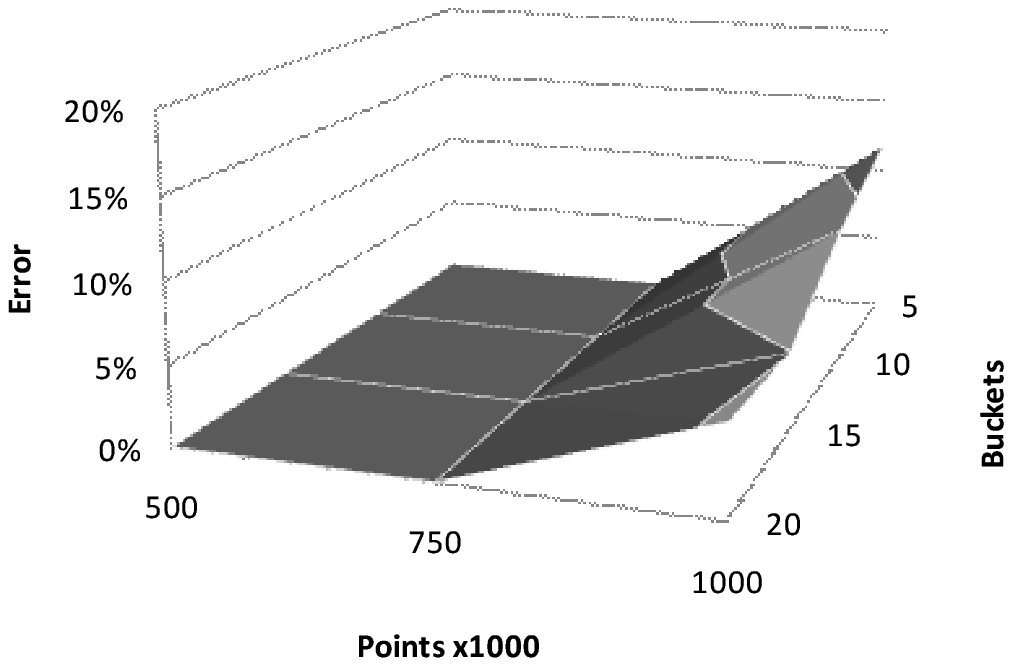}\\
  \caption{{\sc ThresholdRange} excess error, T=100000.}
  \label{fig:TREE100000}
\end{figure}

Figures~\ref{fig:TRE100000} and \ref{fig:TREE100000} show what
happens when we find an interval near the {\sc MaxCount} value. The
two figures show the consequences of the estimation intervals being
offset from the exact intervals by small amounts. The error
decreases with more buckets.

\subsection{\sc ThresholdCount}

This operator is the only operator that does not have relative error
measurements. Instead we report the average number of intervals the
estimation method differs from the exact method. As you can see, we
differ by two from the correct number.

\begin{figure}[h]
  \centering
  \includegraphics[width=4in]{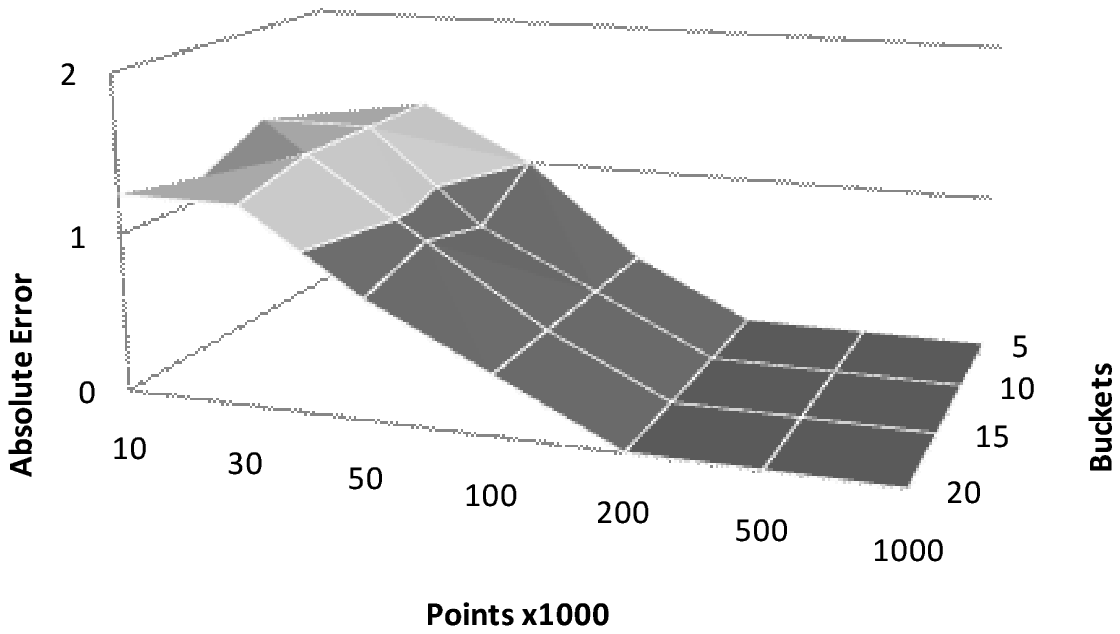}\\
  \caption{{\sc ThresholdCount} error, T=10.}
  \label{fig:TCE10}
\end{figure}

\begin{figure}[h]
  \centering
  \includegraphics[width=4in]{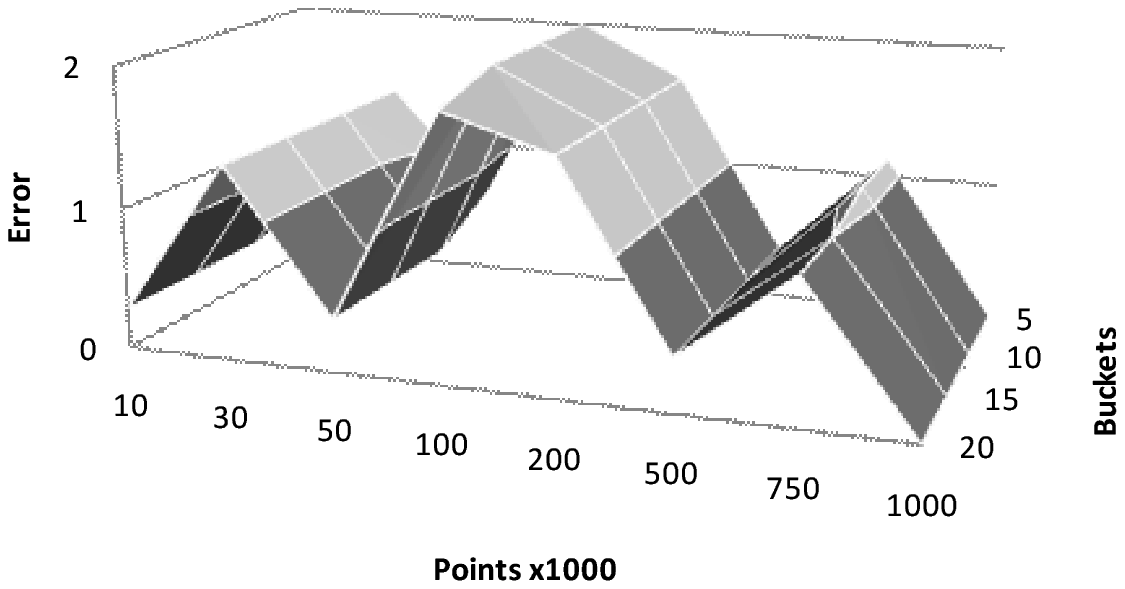}\\
  \caption{{\sc ThresholdCount} error, T=100.}
  \label{fig:TCE1000}
\end{figure}

Figure~\ref{fig:TCE10} shows the average error at $T=10$ where the
errors are small. Figure~\ref{fig:TCE1000} ($T=1000$) looks much
worse, but in reality we are still below 2 intervals off. We also
note that the estimation may split or combine an interval
incorrectly when the intervals are very close together without
greatly affecting the error of other operators. Given this
possibility, the results are excellent.

\subsection{\sc ThresholdSum}

\begin{figure}[h]
  \centering
  \includegraphics[width=4in]{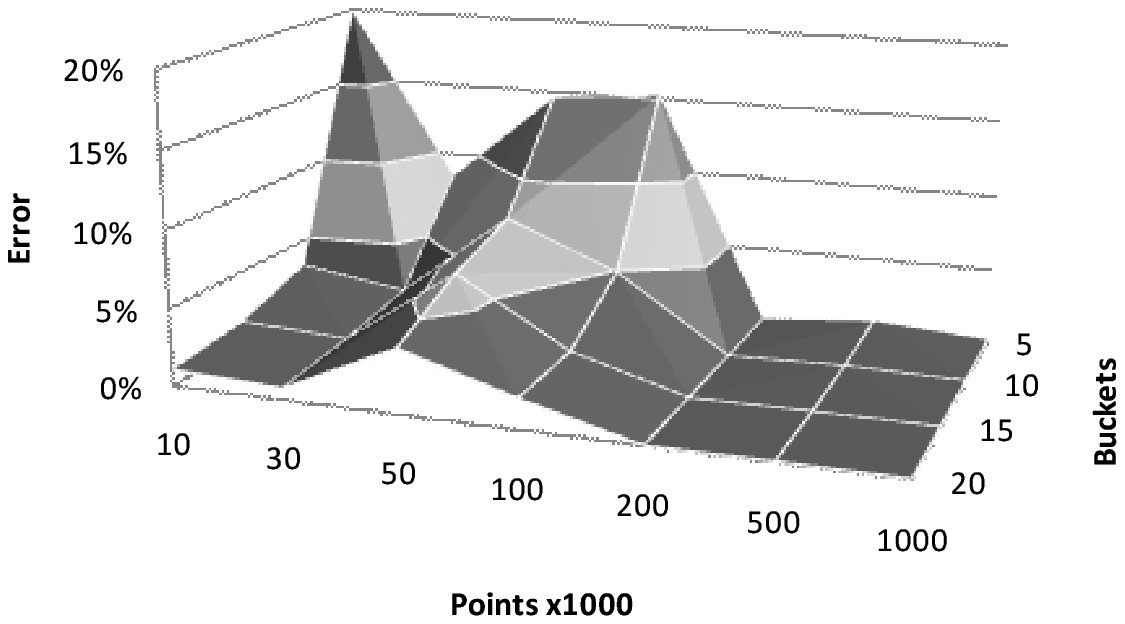}\\
  \caption{{\sc ThresholdSum} error, T=10.}
  \label{fig:TSE10}
\end{figure}

{\sc ThresholdSum} gives the total time above the threshold $T$. As
one can see in Figure~\ref{fig:TSE10}, at higher bucket counts we
have excellent error rates at $T=10$. We didn't always expect great
results at this threshold level across all data sets, but {\sc
ThresholdSum} gives this result consistantly all the way across.

\begin{figure}[h]
  \centering
  \includegraphics[width=4in]{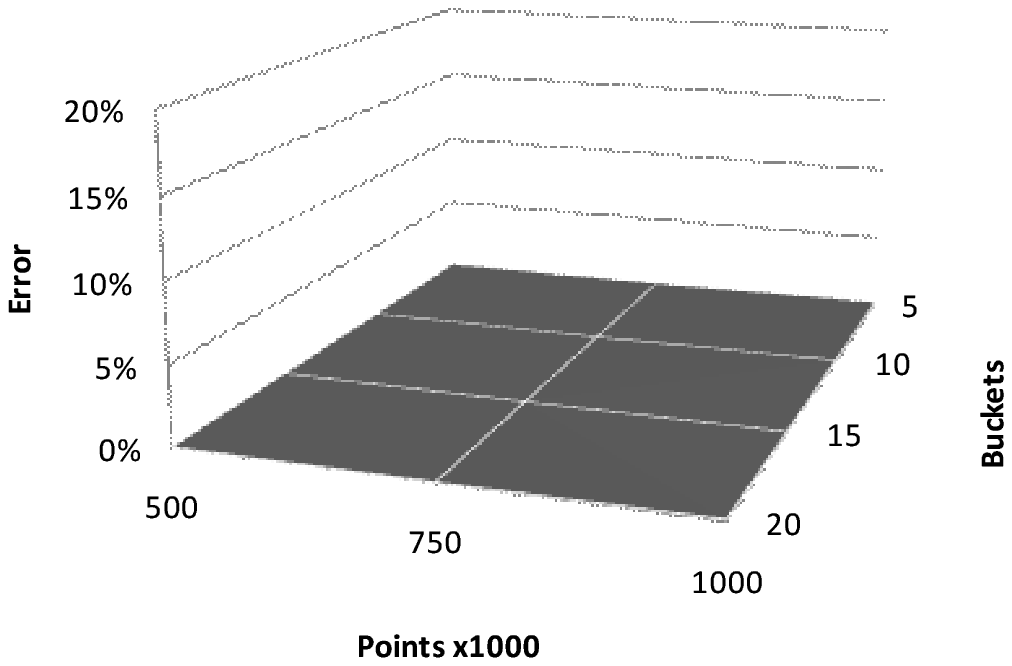}\\
  \caption{{\sc ThresholdSum} error, T=100000.}
  \label{fig:TSE100000}
\end{figure}

We do note that when the threshold approaches {\sc MaxCount}, we see
extremely good accuracy as shown in Figure~\ref{fig:TSE100000}.

\subsection{\sc ThresholdAverage}

\begin{figure}[h]
  \centering
  \includegraphics[width=4in]{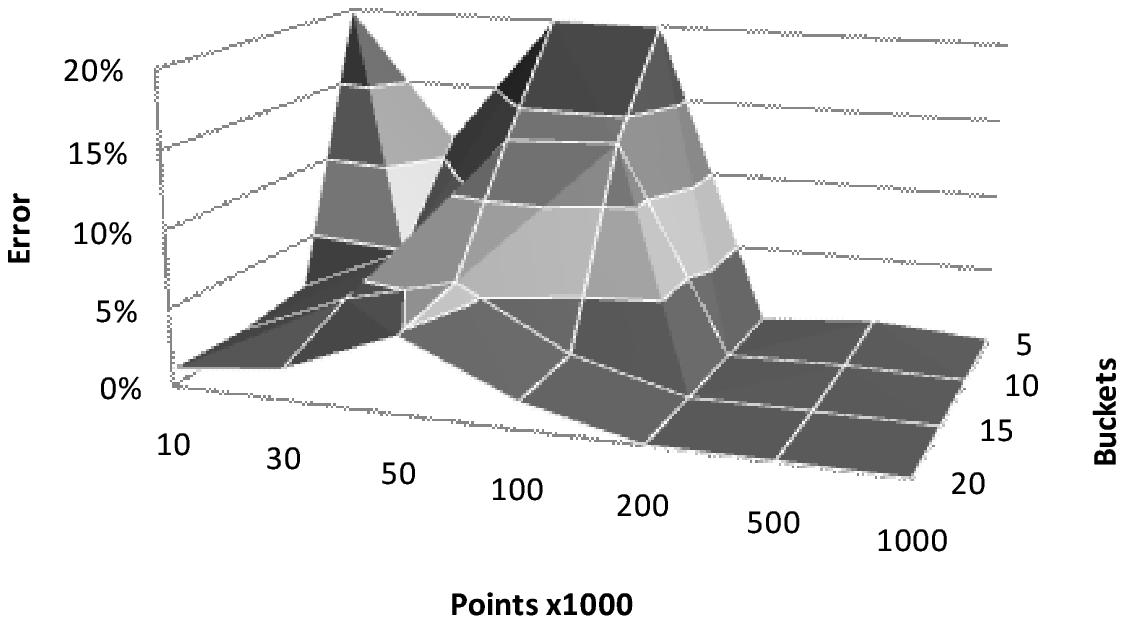}\\
  \caption{{\sc ThresholdAverage} error, T=10.}
  \label{fig:TAE10}
\end{figure}

\begin{figure}[h]
  \centering
  \includegraphics[width=4in]{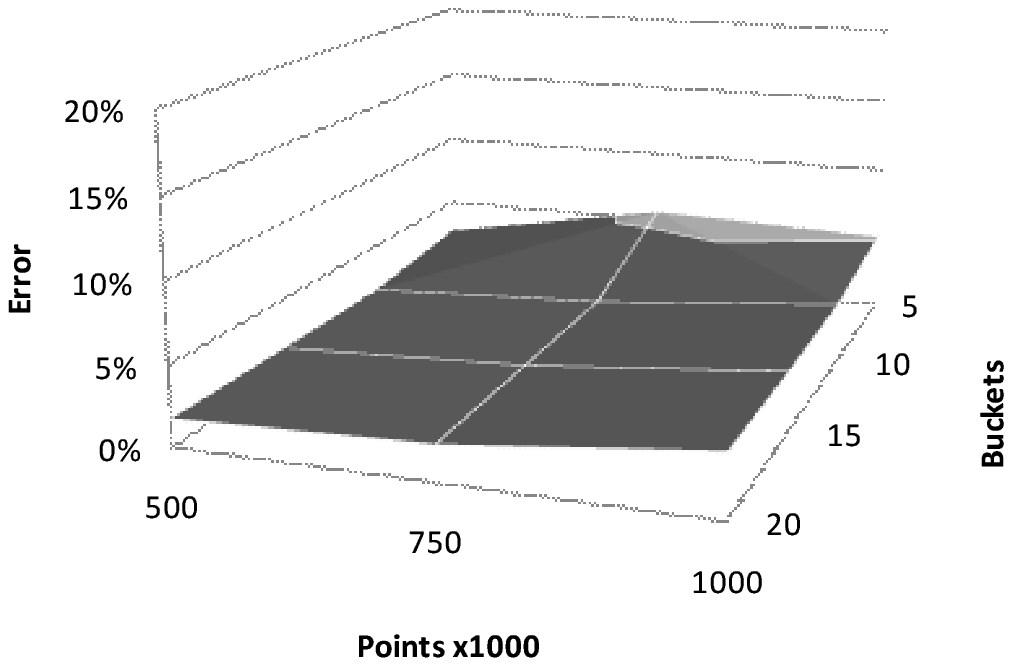}\\
  \caption{{\sc ThresholdAverage} error, T=1000.}
  \label{fig:TAE1000}
\end{figure}

{\sc ThresholdAverage} gives the average length of each time
interval. Figure~\ref{fig:TAE10} shows the now familiar mountains
descending below 5\% error at 20 buckets for $T=10$. The Figure also
shows that even though a few of the data sets tended to have good
results at 5 and 10 buckets, these results are not guaranteed in
general. In Figure~\ref{fig:TAE1000}, the error reaches a plateau
below 5\% with only small bumps in the data.

\subsection{\sc CountRange}

\begin{figure}[h]
  \centering
  \includegraphics[width=4in]{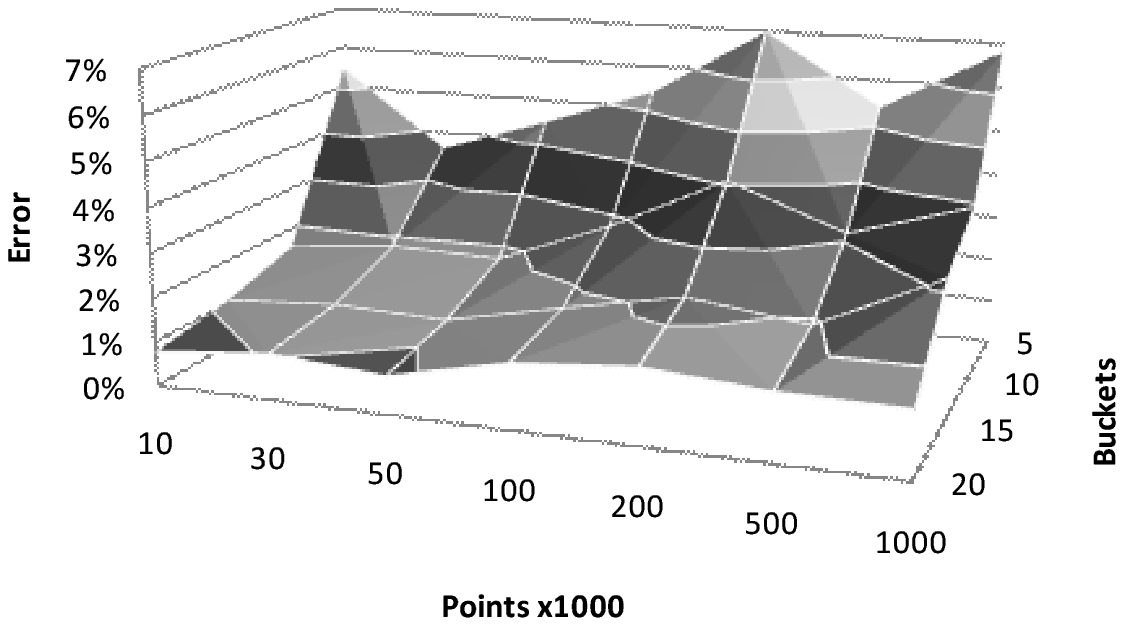}\\
  \caption{{\sc CountRange} error.}
  \label{fig:CountRangeError}
\end{figure}

Other {\sc CountRange} algorithms have achieved error values between
2\% and 3\%. Using our method we conjecture that we could reduce the
error because our method of approximation, although much more
complicated, theoretically adapts to skewed distributions better
than other methods. Figure~\ref{fig:CountRangeError} shows that we
achieved errors under 2\% for 20 buckets across all the data sets,
and in some cases, under 1\%.

Count range also performs about the same speed as the threshold
operators due to its similar implementation.

Additional information that contains error analyses of all the
threshold values is given in~\cite{Anderson2007D}.

\section{Related Work}\label{sec:RelatedWork}

This Section reviews the literature specific to aggregation. Spatial
and spatiotemporal databases have attracted an enormous amount of
interest, and there exists a wide range of literature that is
related to our work only through indexing. For books on the subjects
of spatiotemporal and constraint databases we suggest:
\cite{rigaux2001B,revesz2002,77589,S05Book}, and
\cite{guting_book05}.

\subsection{\textsc{MaxCount} and \textsc{CountRange} Aggregation}

There exists only a few previous algorithms to compute {\sc
MaxCount}~\citep{Revesz20031,Chen20041,Anderson20061}.  None of
those previous algorithms provides efficient queries without
rebuilding the index (i.e., they do not provide dynamic updates).

Previous \emph{approximate} {\sc MaxCount} solutions use indices
from~\cite{APR99} that minimize the skew of point distributions in
the buckets by creating hyper-buckets based on the properties of all
points at index creation time. Updates require the index to be
rebuilt because the buckets depend on the point distribution at a
specific time. In contrast, the probabilistic method we presented
{\em recognizes} point density skew in each bucket instead and
creates a density distribution to model it. We present the first
efficient and dynamic algorithm for {\sc MaxCount}.
Table~\ref{table:results} compares the results of earlier {\sc
MaxCount} algorithms with our current algorithm where $N$ is the
number of points and $B$ is the number of buckets in the index.

\begin{table}[ht]
\centering \caption{{\sc MaxCount} aggregation complexity on
linearly moving objects.} \label{table:results}
\begin{tabular}[bt]{|c|c|c|l|l|l|} \hline
{\bf Max.}& {\bf Worst Case} & {\bf Space}  &  {\bf Exact }     & {\bf Static or}      & {\bf Reference}      \\
{\bf Dim.}& {\bf Time}       &              &  {\bf or Est.}    &
{\bf Dynamic}        & \\ \hline\hline 1         & $O(log\ N)$ &
$O(N^2)$     & Exact             & Static               &
\cite{Revesz20031}   \\ \hline 1         & $O(B \log B)$    & $O(B)$
& Est.              & Static               & \cite{Chen20041}     \\
\hline 2         & $O(B \log B)$    & $O(B)$       & Est. & Static
& \cite{Anderson20061} \\ \hline d         & $O(B)$           &
$O(B)$       & Est.              & Dynamic & \cite{Anderson2007D} \\
\hline d         & $O(N)$           & $O(1)$       & Exact
& Dynamic              & \cite{Anderson2007D} \\ \hline
\end{tabular}
\end{table}

To our knowledge, we present the first proposal of these threshold
aggregate operators for moving points: {\sc MaxCount (and MinCount),
ThresholdRange, ThresholdCount, ThresholdSum}, and {\sc
ThresholdAverage}.

We can modify {\sc Spatiotemporal-Range} algorithms to return the
{\sc CountRange} by counting the objects returned. Several other
algorithms were proposed directly for the {\sc CountRange} problem.
We summarize previous {\sc Spatiotemporal-Range} and {\sc
CountRange} algorithms in Table~\ref{tbl:count}, where $N$ is the
number of moving objects or points in the database, $d$ is the
dimension of the space, and $B$ is the number of buckets. All
algorithms listed are dynamic, which means that they allow
insertions and deletions of moving objects without rebuilding the
index.

\begin{table}[htb]
\centering \caption{{\sc Range} and {\sc CountRange} aggregation
summary.}
\begin{tabular}{|c|l|l|l|l|l|} \hline
{\bf Max.}& {\bf Worst Case}                 & {\bf Worst case}                 & {\bf Exact }       & {\bf Reference}      \\                                         
{\bf Dim.}& {\bf Time}                       & {\bf Space}                      & {\bf or Est.}      & \\ \hline\hline                                                 
2         & $O(N^{\frac{3}{4}+\epsilon}+k)$  & $O(N)$                           & Exact              & \cite{KGT99} \\   
2         & $O(\log_2 N + k)$                & $O(N^2)$\footnotemark            & Exact              &  \\ \hline        
2         & $O(N)$                           & $O(N)$                           & Exact              & \cite{PKGT02} \\ \hline 
3         & $O(N)$                           & $O(N)$                           & Exact              & \cite{SJLL00} \\ \hline 
d         & $O(N)$                           & $O(N)$                           & Exact              & \cite{PLM01} \\ \hline  
d         & $O(B^{d-1} \log_{B}^{d} N)$      & $O(\frac{N}{B}\log_{B}^{d-1} N)$ & Exact              & \cite{ZGTS03} \\ \hline  
2         & $O(\log_B N + C)/B$              & $O(N)$                           & Est.               & \cite{KGT99}\footnotemark \\ \hline                                          
2         & $O(B)$                           & $O(B)$                           & Est.               & \cite{CC02} \\ \hline 
d         & $O(B)$                           & $O(B)$                           & Est.               & Tao et al. (2003) \\ \hline 
d         & $O(\sqrt{N})$                    & $O(N)$                           & Est.               & \cite{TP05} \\ \hline 
d         & $O(B)$                           & $O(B)$                           & Est.               & \cite{Anderson2007D} \\ \hline                                         
\end{tabular}
\label{tbl:count}
\end{table}

\footnotetext[1]{This is a restricted future time query with
expected $O(N)$ space that becomes quadratic if the restriction is
too far into the future.} \footnotetext[2]{$C=K+K'$, where $K'$ is
the approximation error.} \footnotetext[3]{Although \cite{TSP03}
allow dynamic updates, over time the index must be rebuilt.}

In all our work we consider time as a continuous variable. Time as a
discrete variable is discussed in both temporal and spatiotemporal
aggregation by \cite{AAE03,TP05} and \cite{BGJ06}. In the discrete
approach, time stamps describe the temporal nature of objects. This
approach is less relevant to our work, but is relevant to many
applications.

\subsection{Indices and Estimation Techniques}

There are many ways our work is indirectly related to previous work
on indexing structures and estimation techniques. 
{\sc Count} and {\sc Max} aggregation operators have only a titular
relationship to the {\sc MaxCount} aggregation, because one cannot
use the {\sc Count} and {\sc Max} aggregation operators to implement
the {\sc MaxCount} aggregation. Nevertheless, several techniques
used in the {\sc MaxCount} problem are also used in other indices
and algorithms designed for range, max/min, and count queries. We
summarize several of these related techniques next.

\subsubsection{Indices}

The index structure of \cite{AAE03} finds the 2-dimensional moving
points contained in a rectangle in $O(\sqrt {N})$ time.
\cite{GKTD05} gave a selectivity estimation with a histogram
structure of overlapping buckets designed to approximate the density
of multi-dimensional data. The algorithm runs in constant time
$O(d|B|)$, where $d$ is the number of dimensions and $B$ is the
number of buckets. \cite{GKR04} gave a technique for answering
spatiotemporal range, intercept, incidence, and shortest path
queries on objects that move along curves in a planar graph.
\cite{CJNP04,CJSP05} also gave indexing methods that use networks,
such as roads, to predict position and motion changes of objects
that follow roads and characteristics of routes. \cite{PJ05} used
networks to reduce the dimensionality of constrained moving object
to two dimensional trajectories and examined the method in terms of
the spatiotemporal range query. \cite{AG051} proposed the MON-Tree
to index moving objects in networks using graphs or route oriented
networks to find the spatiotemporal range and windows queries. They
define window queries as returning the pieces of the object's
movement that intersects the query window. \cite{ZMTGS01} proposed
the {\em multiversion SB-tree} to perform range temporal aggregates:
{\sc Sum}, {\sc Count} and {\sc Avg} in $O(\log_b n)$, where $b$ is
the number of records per block and $n$ is the number of entries in
the database. \cite{TIME05_Revesz} gave efficient rectangle indexing
algorithms based on point dominance to find count interpreted in $k$
dimensions using the following concepts:
\begin{enumerate}
\item {\em stabbing} gives the number of objects that contain a point;
\item {\em contain} gives the number of rectangles that contain the query rectangle;
\item {\em overlap} gives the number of rectangles that overlap the query rectangle; and
\item {\em within} gives the number of rectangles within the query space.
\end{enumerate}
These four operators have a running time of $O(\log^k n)$ where $k$
is the number of dimensions and $n$ is the number of points.

\cite{SJLL00} gave an R$^{\ast}$-tree based indexing technique for
1, 2, and 3 dimensional moving objects that provide time-slice
queries (selection queries), windows queries, and moving queries.
Window queries return the same information as range queries, but
with a valid time window starting at the current time and continuing
to $t_h$. Window queries may request predictions for range queries
within this window of time. Moving queries, similar to incidence
queries, return the points that are contained within the space
connecting one rectangle at a start time to a second rectangle at an
end time. The proposed time parameterized R-tree (TPR-Tree) search
runs in expected {\em logarithmic time}. Another R$^{\ast}$-tree
extension given by \cite{CR00} forms tighter parametric bounding
boxes than~\cite{SJLL00} and has similar running time. \cite{Tao03}
proposed the TPR$^\ast$-Tree that extends the TPR-Tree with improved
insert and delete algorithms. In the context of a variety of count
queries it performs similarly to previous indices.

\cite{SPTL04} uses time-dependent, updatable, histograms to query
counts at specific times including past, present and future.
Recently, \cite{PSJ06} proposed the $R^{PPF}$-tree that indexes
past, present and predictive positions of moving points, and extends
the previous work on TPR-Trees \citep{SJLL00} with a partial
persistence framework. Earlier work by \cite{TayebUW98} adapted the
PMR-quadtree~\citep{77589}, a variant of the quadtree structure, for
indexing moving objects to answer time-slice queries, which they
called instantaneous queries, and infinitely repeated time-slice
queries, called continuous queries. Search performance is similar to
quadtrees and allows searches in $O(\log N)$ time.

\cite{MSI02} use the sweeping technique from computational geometry
to define a query language to evaluate past, present, and future
positions of moving objects in constraint databases.

Finally, \cite{HadjieleftheriouKGT03} use an efficient approximation
method to find areas where the density of objects is above a
specific threshold during a specific time interval. This method
comes the closest to the method used in our aggregation operators,
but does not allow for the query to move or change shape over time.
In fact, this method is not applied to counting at all.

Note that each of these indexing methods that return the moving
points in a query window or rectangle can be easily modified to
return instead the {\em count} of the number of moving points.
However, they may not be easily extended to provide a {\sc MaxCount}
within a changing, moving query space.

With a few exceptions you can see that {\sc Count} aggregation is
$O(\log N + d)$ for exact methods and $O(B)$ or better for
estimation methods. The hidden constant in the exact method is the
number of buckets that must be traversed to find the {\sc Count}.
Estimation methods vary in many ways and asymptotic running time
doesn't always give a meaningful estimate as to how big $B$ will be.


\subsubsection{Estimation Techniques}

Our work is related to several other papers that {\em estimate} the
count aggregate operation on spatiotemporal databases.

\cite{APR99} gave an algorithm that can estimate the {\sc Count} of
the number of the rectangles that intersect a query rectangle for
Selectivity Estimation. \cite{CC02} and \cite{TSP03} proposed
methods that can estimate the {\sc Count} of the moving points in
the plane that intersect a query rectangle. More recently,
\cite{KPGT05} gave a predictive method based on dual
transformations.

\cite{WolfsonY03} and \cite{TWHC04} gave a method for generating
pseudo trajectories of moving objects. Most of these estimation
algorithms use {\em buckets} as basic building structures of the
index.  In extending this idea, we use $2d$-dimensional
hyper-buckets in our algorithms where $d$ is the number of
dimensions in the moving-objects space.

\section{Conclusions and Future Work}\label{sec:Conclusions}

We implemented and compared two new {\sc MaxCount} algorithms. The
estimated {\sc MaxCount} was shown to be fast and accurate while
still allowing fast constant time updates. No other algorithm has
these features to date. We showed that {\sc ThresholdRange}, {\sc
ThresholdCount}, {\sc ThresholdSum}, {\sc ThresholdAverage}, and
{\sc CountRange} are related to {\sc MaxCount} and can be evaluated
using similar techniques and that we achieve error values under 5\%
in these operations. We gave an empirical threshold for choosing
between the exact and estimated algorithms. We discussed the issues
related to higher dimensions and note that all sweeping algorithms
have this problem. We also note that using our technique it is
possible to decompose the problem and run it in a multiprocessor or
grid environment where the database is divided into smaller
databases.

Future work may include decreasing the running time by finding other
techniques because there does not appear to be a clear method for
decreasing the running time of sweeping methods. One could also
consider implementing and comparing these techniques in a grid
computing environment.

\bibliographystyle{agsm}
\bibliography{h:/unl/BibTex/all}

\end{document}